\documentclass[aps,prl,reprint,superscriptaddress]{revtex4-2}

\usepackage{braket}
\usepackage{quantikz}
\usepackage{graphicx}
\usepackage{dcolumn}
\usepackage{bm}
\usepackage{lipsum}
\usepackage{amsthm}
\usepackage[colorlinks,linkcolor=blue,urlcolor=blue, citecolor=blue]{hyperref}
\usepackage[normalem]{ulem} 

\newcommand{\ketbra}[2]{|#1\rangle\! \langle #2|}

\def\captionof#1#2{{\def\@captype{#1}#2}}

\newtheorem{theorem}{Theorem}[]

\begin{document}

\preprint{APS/123-QED}

\title{The Impact of Imperfect Timekeeping on Quantum Control}

\author{Jake Xuereb}
\affiliation{Vienna Center for Quantum Science and Technology, Atominstitut, TU Wien, 1020 Vienna, Austria}
\author{Florian Meier}
\affiliation{Vienna Center for Quantum Science and Technology, Atominstitut, TU Wien, 1020 Vienna, Austria}
\author{Paul Erker}
\affiliation{Vienna Center for Quantum Science and Technology, Atominstitut, TU Wien, 1020 Vienna, Austria}
\affiliation{Institute for Quantum Optics and Quantum Information (IQOQI), Austrian Academy of Sciences,
Boltzmanngasse 3, 1090 Vienna, Austria}
\author{Mark T. Mitchison}
\affiliation{School of Physics, Trinity College Dublin, College Green, Dublin 2, Ireland}
\affiliation{Trinity Quantum Alliance, Unit 16, Trinity Technology and Enterprise Centre, Pearse Street, Dublin 2, D02YN67}
\author{Marcus Huber}
\affiliation{Vienna Center for Quantum Science and Technology, Atominstitut, TU Wien, 1020 Vienna, Austria}
\affiliation{Institute for Quantum Optics and Quantum Information (IQOQI), Austrian Academy of Sciences,
Boltzmanngasse 3, 1090 Vienna, Austria}

\date{\today}

\begin{abstract}
In order to unitarily evolve a quantum system, an agent requires knowledge of time, a parameter which no physical clock can ever perfectly characterise. In this letter, we study how limitations on acquiring knowledge of time impact controlled quantum operations in different paradigms. We show that the quality of timekeeping an agent has access to limits the circuit complexity they are able to achieve within circuit-based quantum computation. We do this by deriving an upper bound on the average gate fidelity achievable under imperfect timekeeping for a general class of random circuits. Another area where quantum control is relevant is quantum thermodynamics. In that context, we show that cooling a qubit can be achieved using a timer of arbitrary quality for control: timekeeping error only impacts the rate of cooling and not the achievable temperature. Our analysis combines techniques from the study of autonomous quantum clocks and the theory of quantum channels to understand the effect of imperfect timekeeping on controlled quantum dynamics.
\end{abstract}

\maketitle


Quantum control theory is an established field~\cite{Wiseman2009,Cong2014} with notable successes~\cite{Koch2022}, yet its basic formulation implicitly assumes that the controlling agent has perfect knowledge of time: an assumption which is never truly satisfied. Indeed, a long-standing problem in quantum theory has been to understand how resource limitations fundamentally restrict our knowledge of time~\cite{Salecker1958, peres, thermal_clocks, Woods2019,Woods2021autonomousticking,manu}. Peres closes his seminal paper \cite{peres} with the warning ``\textit{\dots the Hamiltonian approach to quantum physics carries the seeds of its own demise}'', lamenting his conclusion that controlled Hamiltonian dynamics is unachievable without access to a perfect clock requiring infinite energy. In this work, we examine whether a central task in quantum theory, that of controlling unitary operations, is limited by one's ability to tell time. In doing so, we give credence to Peres' admonition but also provide justification for the timescales over which it is fair to ignore it.

In classical computation, bits of information are abundant and their manipulation is practically error free. This underpins the great success of the abstract theory of classical information processing, which allows the design of algorithms without much consideration for the physical processes that carry them out. Quantum computation is different. Whilst any algorithm before measurement can simply be thought of as a rotation on a high-dimensional Bloch ball, such a unitary operation is necessarily generated by a Hamiltonian. This governs a physical process which is sensitive to physical control parameters whose finite precision leads to error in the unitary it produces---in this work we focus on time.

Characterising error, its propagation and impact is a well studied problem in quantum computation, especially within the field of randomized benchmarking~\cite{Emerson_2005,Magesan_2011,Carignan_Dugas_2018}. This characterisation has been successful in its generality because it is agnostic to the physical source creating them. In contrast, we focus on studying the relationship between one particular error and the physical source generating it. Specifically, we explore the error contributed by the imperfect nature of the timekeeping device that regulates the physical control protocol, neglecting all other error sources. Our central motivation is to examine what quantum computational tasks are achievable depending on the quality of the clock one has access to. The role of timekeeping in quantum computation has been experimentally \cite{Ball2016TheRO,fast_quantum_gates,he2022quantum} and theoretically \cite{jiang_sensitivity_laser} investigated before, even being accounted for in experimental quantum computing setups~\cite{ballance,ringbauer_qudit}. However, a general understanding of how timekeeping errors accumulate according to the structure of a circuit is still lacking. 

In this work, we address this gap by establishing a rigorous connection between the fidelity of a quantum process and the quality of available timekeeping devices, exploiting a framework developed in the emerging field of autonomous quantum clocks \cite{thermal_clocks,Woods2019,manu,Woods2021autonomousticking} known as the tick distribution. A key property of a clock's tick distribution is its accuracy $N$, which is the expected number of ``good'' ticks before the clock becomes inaccurate (see Eq.~\eqref{accuracy} below for a precise definition). We show that incorporating a tick distribution within unitary time evolution enacts a dephasing channel, empowering us to use techniques for characterising noisy channels to quantify the impact of imperfect timekeeping on quantum operations.

The main contributions we present in this letter are: firstly, that the average gate fidelity of ill-timed single and two qubit gates obey the fidelity relation
\begin{gather*}\overline{\mathcal{F}} = \frac{2 + e^{-\frac{\theta^2}{2 N}}}{3},\end{gather*}
where input states non-trivially rotated by the gate are exponentially impacted by the accuracy $N$ of the timer and $\theta$ is the pulse area (i.e.~the angle of rotation on an effective Bloch sphere) due to the gate. We make use of this insight to upper bound the average gate fidelity achievable under imperfect timekeeping in generic circuits. Secondly, we show that with arbitrarily imperfect timekeeping one can still cool a qubit to a desired temperature with access to asymptotically many ill-timed SWAP operations and a thermal machine. Throughout, we set $\hbar=1$ throughout.

\textit{Imperfect timekeeping in quantum control.---}An ideal unitary quantum operation $U=e^{-iH\tau}$ requires the ability to perfectly control the duration $\tau$ for which we allow a physical process, generated by a Hamiltonian $H$, to act on a target system. One can think of this as having access to a timer which has infinite accuracy and precision in its ability to tick at a desired time and using it to control the duration of this process. In this context, imperfect timekeeping is the situation where one has access to an imperfect timer which does not always tick at the desired time to end this operation and so can result in a different unitary evolution than one expects.

We may model the quality of our timekeeping in this control process by introducing a probability distribution obtained by sampling from the timer which we are using for control. Such distributions are known as tick distributions and have been studied within the field of quantum clocks \cite{thermal_clocks,Woods2019,Woods2021autonomousticking,manu}. If the timer is perfect then we always obtain the desired duration to time our process, leading to an evolution which gives the target state
\begin{gather}
    \rho' = \int^{\infty}_{-\infty}dt\, \delta\left(t-\tau\right)e^{-iHt}\rho e^{iHt},
\end{gather}
where $\rho$ is the initial state and the Dirac delta function $\delta(t-\tau)$ represents the tick distribution of an ideal timer. Instead, let us make this scenario more realistic by considering a Gaussian tick distribution, giving on average a final state of the form 
\begin{gather}\label{eq:gaussian_err_state}
\rho'_{\text{error}} = \int^{\infty}_{-\infty} dt\, \frac{e^{\frac{(t-\tau)^2}{-2\sigma^2}}}{\sqrt{2\pi\sigma^2}} e^{-iHt}\rho e^{iHt} ,
\end{gather}
where $\sigma$ is the variance of the tick distribution and our desired process duration $\tau$ is its mean. The accuracy, defined by~\cite{thermal_clocks} 
\begin{gather}
\label{accuracy}
    N = \left(\frac{\tau^2}{\sigma^2}\right),
\end{gather}
is a figure of merit ascribed to a clock which generates ticks with a temporal resolution $\tau$ and variance $\sigma^2$. In this work, we focus on single-tick timers which control a gate pulse, and so the relevant clock accuracy is defined with respect to the resolution of the pulse duration, $\tau$.

The states $\rho'$ and $\rho'_{\text{error}}$ are clearly not the same in general. Let the Hamiltonian be of form $H = \sum_{n} E_n \ketbra{n}{n}$ so that we may express the contributions to the resultant state in the energy eigenbasis as
\begin{align}
\bra{n} \rho'_{\text{error}} \ket{m} &= \int^{\infty}_{-\infty}dt\, \frac{e^{\frac{(t-\tau)^2}{-2\sigma^2}}}{\sqrt{2\pi\sigma^2}}\bra{n} e^{-iHt}\rho e^{iHt}\ket{m}\notag\\
&=\int^{\infty}_{-\infty}dt\frac{e^{\frac{(t-\tau)^2}{-2\sigma^2}}}{\sqrt{2\pi\sigma^2}}e^{-i\left(E_n - E_m\right)t}\bra{n} \rho \ket{m} \notag \\
& = e^{-\frac{\sigma^2}{2}\left( E_n - E_m\right)^2} e^{-i\left(E_n-E_m \right)\tau}\bra{n} \rho \ket{m}\label{eq:gaussian_dephasing}
\end{align}
This admits two cases. For the diagonal entries $m=n$, the unitary has no impact on the state in the basis of its Hamiltonian generator as can be seen from \eqref{eq:gaussian_dephasing}. For the off-diagonal elements with $m \neq  n$ we see that the term $e^{-i\left(E_m-E_n\right)\tau}$ enacts the unitary evolution in the basis of the gate Hamiltonian and the term $e^{-\frac{\sigma^2}{2}\left(E_m - E_n\right)^2}$ causes decoherence due to the timing error. Therefore, imperfect timekeeping is described by a dephasing map $\mathcal{D}(\cdot)$ over the unitary operation we are trying to carry out:
\begin{gather}
    \rho'_{\text{error}} = \mathcal{D}\left(U\rho U^\dagger\right), \label{eq:dephasing}
\end{gather}
where $\mathcal{D}(\cdot)$ denotes a dephasing channel in the eigenbasis of the Hamiltonian which generates $U$.

In the Appendices we show that Eq.~\eqref{eq:dephasing} holds for arbitrary (non-Gaussian) tick distributions. However, in general we find that the variance alone is not sufficient to characterize the channel: the rate of dephasing depends on the full characteristic function of the tick distribution. Nevertheless, is worth pointing out that most physical timers would have a tick distribution which can be considered Gaussian. This is a result of the central limit theorem~\cite{Klenke2020} if the timer obtains $\tau$ by integrating over sufficiently many shorter ticks. Then, $\tau$ is Gaussian distributed in leading order of the number of ticks over which is summed and as such, choosing the Gaussian regime within this analysis is not overly restrictive.

\begin{figure*}[ht!]
    \centering
\includegraphics[width = \textwidth]{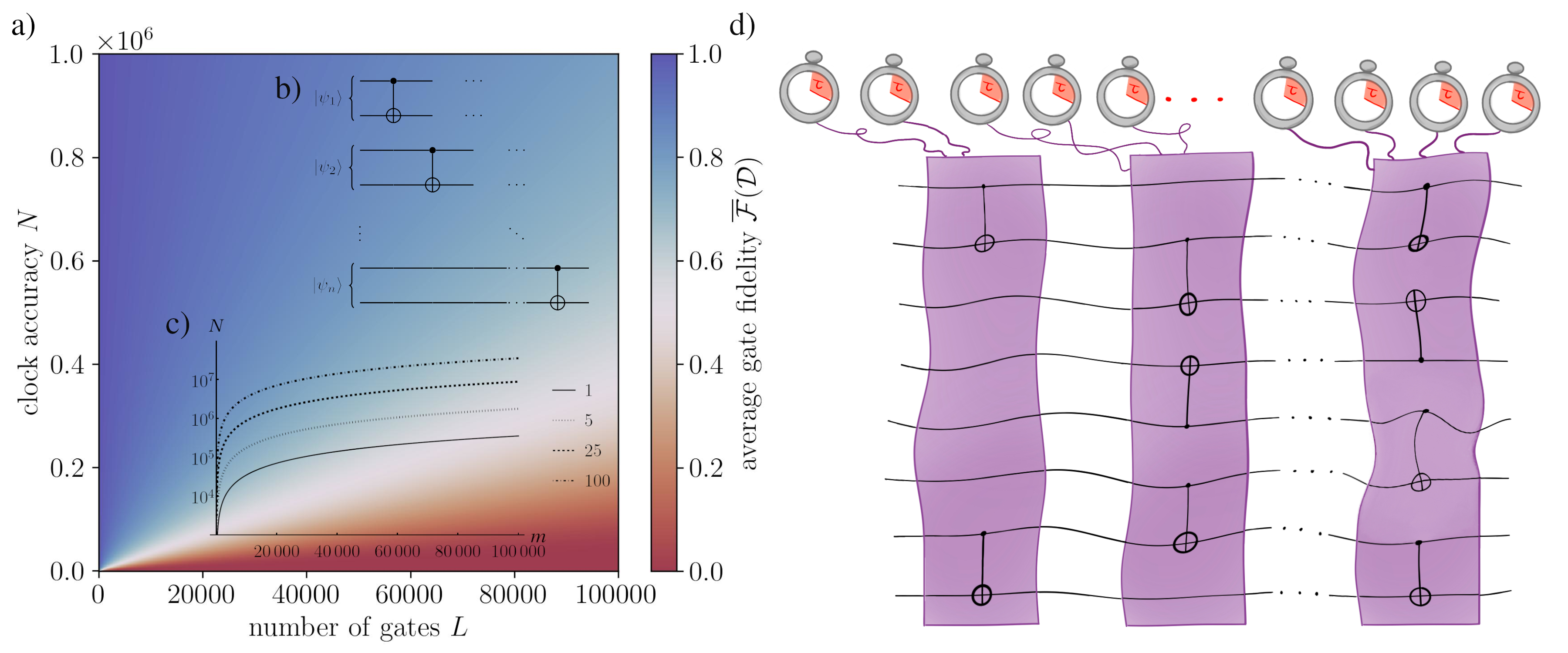}
\caption{For a circuit comprised of $l_t = 1$ CNOT per timestep as the one in b), the impact of timekeeping error on average gate fidelity $\overline{\mathcal{F}}(\mathcal{D})$ is plotted in a) as the number of gates $L$ increases for different values of clock accuracy $N$. Inset c) is the clock accuracy in logarithmic scaling against the depth $m$ for different no. of CNOTs per timestep i.e. $l_t = \{1, 5, 25, 100\}$, given a threshold average gate fidelity of 0.5~\cite{footnote_fidelity}. This shows that the temporal control one has access to greatly impacts the circuit complexity one can achieve and that this relationship is not linear. In the right panel d), we present an illustration which visually conveys the setting of Theorem~1. That is, $l_t$ CNOTs per step of an algorithm individually timed by independent clocks following identical tick distributions leading to independent dephasing at each time step. This results in a global dephasing map whose unitarity we use to bound the achievable average gate fidelity.\label{fig:cnot}}
\end{figure*}

\textit{Fidelity of ill-timed quantum logic gates.---}The dephasing of a single qubit in a given basis is described by a channel whose Kraus operators have the form~\cite{petruccione_breuer,Bylicka2014NonMarkovianityAR} 
\begin{align}
    K_1 = \sqrt{\frac{1 + e^{-\Gamma}}{2}}I && K_2 =  \sqrt{\frac{1 - e^{-\Gamma}}{2}}\sigma_z,
\end{align}
where $\Gamma$ is the dephasing magnitude, $I = \ketbra{0}{0} + \ketbra{1}{1}$ is the identity operator and $\sigma_z = \ketbra{0}{0} - \ketbra{1}{1}$ is the operator corresponding to the population difference in the chosen basis. In the context of Eq.~\eqref{eq:gaussian_dephasing}, the states $|0\rangle$ and $|1\rangle$ are the eigenstates of the gate Hamiltonian, $H$, while $\Gamma = \sigma^2\Omega^2/2$ with $\Omega = E_0 - E_1$ is the Rabi frequency induced on the system by the control field which mediates the gate interaction.

The average gate fidelity is a measure of how much degradation a channel causes on the unitary evolution of a Haar-random qubit state for a given gate and as such presents an ideal candidate for examining the impact of imprecise timekeeping on a quantum computation. It was shown in~\cite{Nielsen_2002,Emerson_2005} that the average gate fidelity is dependent only on the noisy channel $\mathcal{E}(\cdot) = \sum_i K_i (\cdot)K_i^\dagger$ being considered as $
\overline{\mathcal{F}}(\mathcal{E}) = \frac{\sum_i |\text{tr} K_i|^2 + d}{d^2 +d},$
where $d$ is the dimension of the Hilbert space we are averaging over. Applying this expression to our dephasing channel we obtain the average gate fidelity
\begin{gather}
\overline{\mathcal{F}}(\mathcal{D}) = \frac{2 + e^{-\frac{\theta^2}{2N}}}{3}, \label{eq:infidelity}
\end{gather}
expressed in terms of the pulse area $\theta =\Omega\tau$ and the accuracy defined in Eq.~\eqref{accuracy}. Equation~\eqref{eq:infidelity} exposes the exponential impact of clock accuracy on gate fidelity where $2/3$ appears as a contribution from states not rotated by the gate and thus unaffected by the error. Note that this impact is dependent on the size of the rotation on the Bloch ball related to the gate, such that larger rotations are more susceptible to being impacted by the quality of timekeeping.

We can extend this result to the CNOT gate which,  when combined with single qubit rotations, forms a universal gate set~\cite{nielsen_chuang_2010}. The CNOT can be generated by a Hamiltonian that only acts on the subspace spanned by $\ket{10}$, $\ket{11}$ allowing us to consider this operation as acting on an effective qubit with basis states $\ket{10}$,$\ket{11}$. The impact of imperfect timekeeping will be dephasing within this non-local subspace. The resultant average gate fidelity of ill-timed two qubit entangling gates is therefore given by Eq.~(\ref{eq:infidelity}) with $\theta = \pi$. This generalisation is explored in further detail in the Appendices.

As a measure of the complexity of a quantum circuit in this work we will consider the CNOT count due to its theoretical relevance in constructing minimal gate decompositions~\cite{Amy_2018,CNOT_count_3,CNOT_count_4} and practical relevance in design of quantum computing architectures~\cite{CNOT_count_2}. Additionally, one would expect that non-local errors emerging from imperfectly timed entangling gates are harder to correct for when compared to local errors stemming from ill-timed local gates. So focusing on the impact of imperfectly timed entangling gates e.g. CNOT, on the achievable gate fidelity of a circuit gives a reasonable upper bound on achievable fidelities of the total circuit. It is interesting to then ask what circuit complexity is achievable with access to a given quality of timekeeping resource. This question is not straightforward to investigate as the impact of timekeeping error on a circuit is highly dependent on its structure and input state. As an initial inquiry, we obtain a result for the impact of imperfect timekeeping on generic circuits involving independently and identically timed CNOTs which do not intersect within a given step, but which can overlap between different steps.
\begin{theorem}
The average gate fidelity of a quantum circuit on $n$ qubits with depth $m$ and $l_t$ CNOTs applied to some $2l_t \leq n$ distinct pairs of qubits at each time step $t$ is upper bounded
\begin{gather}
\overline{\mathcal{F}} \leq \frac{2^{n}\left(\frac{1 + e^{\frac{-\pi^2}{N}}}{2}\right)^{\frac{L}{2}}+1}{2^{n}+1}, \label{eq:scaling}
\end{gather}
if each CNOT is timed by an independent clock with an identical Gaussian tick distribution with mean $\pi$ and variance $\sigma$. Where $L = \sum^m_{t=1} l_t$ is the total number of CNOTs in the circuit and $N$ is the clock accuracy.
\end{theorem}

\textit{Overview of proof.---}This theorem is derived using the observation that imperfect timekeeping is a dephasing map and that each gate is timed by independent clocks with identical tick distributions. In a given timestep $t$ this results in $l_t$ non-local single qubit subspaces being independently dephased, each via the Kraus operators $K_1,K_2$ giving a large dephasing map with Kraus operators $\Lambda_{t}$ produced by $2^{l_t}$ tensor products of $K_1$ and $K_2$. For the total circuit we have a concatenation of $m$ such large dephasing maps which gives one global dephasing map whose average gate fidelity we seek to upper bound. The average gate fidelity of a concatenation of \textit{non-destructive} channels, e.g., dephasing channels, is known to be upper bounded~\cite{dugas_unitarity_bound,Carignan_Dugas_2019} by the unitarity $\Upsilon$ as $\overline{\mathcal{F}} \leq \frac{d \Upsilon + 1}{d+1}$ where $\Upsilon = \sum_t \frac{||\Lambda_t||^2_2}{d}$, $||\cdot||_2 = \sqrt{\text{tr}\{\cdot^\dagger\cdot\}}$ is the Schatten 2-norm and $d$ is the dimension of the Kraus operator. Since this is a 2-norm and the Pauli operators are involutary, $\Upsilon$ will be comprised of products of the scalar factors of $K_1$ and $K_2$ since
$\Lambda_{t}^{\dagger} \Lambda_{t} = \left(\frac{1+e^{-\Gamma}}{2}\right)^{l_t -i}\left(\frac{1-e^{-\Gamma}}{2}\right)^{i} I_n$
for $i \in [0,l_t]$ in a given timestep. Considering products of such terms for the Kraus operators of the depth $m$ concatenation gives $\Upsilon = \left(\frac{1 + e^{\frac{-\pi^2}{N}}}{2}\right)^{L/2}$ and so the theorem. Further details of this proof are given in the Appendix.

Whilst explicitly framed around CNOTs, as a construction this theorem could be applied to any Clifford+T circuit~\cite{boykin,matsumoto2008representation} which can be split into blocks or steps which internally commute but where individual blocks do not commute with each other. This gives us an insight into how one should expect imperfect timekeeping to impact quite generic quantum algorithms as they scale. For perspective, we can consider some rough estimates. Setting an average gate fidelity threshold at 0.5 for the entire computation, Eq.~\eqref{eq:scaling} for 1 CNOT per time step suggests that we can execute a circuit such as Fig~\ref{fig:cnot} b) featuring $m=1\times\,10^6$ CNOTs, given a clock accuracy of $N=3.6\times\,10^6$. At the same fidelity threshold, $m=1\times\,10^4$ CNOTs are achievable with a clock accuracy of $N=3.6\times\,10^4$, as shown in Fig.~\ref{fig:cnot} a). This circuit complexity is consistent with state-of-the-art benchmarking algorithms, where CNOT gates may number in the tens of thousands~\cite{Amy_2018}. Considering a gate duration of $\tau = 100$ ns, these accuracies would correspond to timing uncertainties of $\sigma \approx 0.168$ ns and $\sigma \approx 0.530$ ns, respectively. Modern control systems such as ARTIQ and Sinara suffer from electronic jitter of 1--0.1~ns~\cite{artiq_sinara}, signalling that timekeeping is still a limiting factor for experiments. However, typical experimental error budgets~\cite{fast_quantum_gates,ringbauer_qudit} include other timekeeping errors beyond jitter, including the frequency instability of local oscillators~\cite{Ball2016TheRO} and peak disortion in their signal. This leads us to believe that the identification and quantification of all sources contributing to experimental timekeeping uncertainty remains an important task. Doing so would help to connect the physics of quantum control with recent theoretical progress in the foundations of timekeeping~\cite{Woods2019,thermal_clocks,Woods2021autonomousticking,manu,he2022quantum}.

\textit{Imperfect timekeeping is sufficient for cooling.---}Our results so far corroborate an accepted wisdom in the practice of quantum computation: timing operations precisely and fast is a key technological challenge that will have to be met for fault-tolerant quantum computation~\cite{aharonov1997fault,bravyi_kitaev_fault,Knill_1998}. This being said, unitary operations generated by controlled Hamiltonians also appear in more conceptual areas of quantum theory, e.g.~in quantum thermodynamics. In that context, the ramifications of imperfect timekeeping have only recently begun to come to light~\cite{Malabarba2015,WoodsHorodecki2019,Taranto2021}. As a concrete thermodynamic protocol, we consider algorithmic cooling~\cite{park2016heat,imperfect_thermalisation,Clivaz_2019_cost,Clivaz_2019_bound}, where a series of unitary operations are used to perform refrigeration.  Such protocols can be broken down into SWAP-inducing interactions enacted between the energy levels of a thermal machine and the system to be cooled. In \cite{Silva_2016}, the achievability of this cooling protocol is shown for the task of cooling a qubit to any desired temperature given access to an appropriate thermal machine, whilst bounds on its performance were obtained in~\cite{Clivaz_2019_bound}.
\begin{theorem}
A qubit in a thermal state at temperature $\beta_s$ with ground state population $r_s$ can be asymptotically cooled to a thermal state with ground state population $r_v \, : \, 1\geq r_v > r_s\geq 0$ and inverse temperature $\beta_v \, : \, \infty>\beta_v > \beta_s \geq 0$ in a protocol controlled by a timer with a Gaussian tick distribution with mean $\tau > 0$ and variance $\sigma >0 $.
\end{theorem}
Here is a sketch of the proof. A pair of energy levels of the thermal machine are chosen, such that the subspace they form can be thought of as a virtual qubit with inverse temperature $\beta_v \, : \, \beta_v > \beta_s$. The probability that this subspace is populated is $P_v: 0<P_v<1$, which is less than unity because the machine may have many levels. To cool the physical qubit, the agent attempts to generate a SWAP operation between the virtual and physical qubits by enacting a Hamiltonian for a time $\tau$ using a timer with finite accuracy $N$. The imperfect timer dephases this operation, resulting in the ground state population of the virtual qubit being only partially swapped with that of the physical qubit. Attempting to apply $n$ SWAPs recursively, and assuming the thermal machine can fully relax at each step, the agent manages to change the ground state population of the physical qubit to
\begin{gather}
 r_{\text{error}}^{(n)} = r_v - (r_v - r_s)(1 -P_v(1-p))^n,
\end{gather}
where $p = \tfrac{1}{2} \left(1 - e^{-\pi^2/2N} \right)$ appears due to the imperfect temporal control. Asymptotically as $n \to \infty$, $ r_{\text{error}}^{(n)}$ converges to $r_v$, i.e.\ the physical qubit converges to a thermal state with the target temperature $\beta_v$, regardless of the control timer's uncertainty $\sigma$. Whilst imperfect timekeeping is sufficient for an agent to cool a qubit to a desired temperature, there is a resource cost in terms of the number of SWAP operations required, which scales inversely with the accuracy of the timer. We examine how $\sigma$ impacts the rate of cooling in this protocol in~\cite{sup_mat}. 

\textit{Discussion.---}In this work we have shown that Peres' concern is not only relevant to the foundations of quantum theory but also to operational tasks such as computation and refrigeration. Within the context of our model, Figure~\ref{fig:cnot} shows that quantum algorithms with different circuit complexity are achievable depending on clock accuracy and that this relationship is nonlinear. The clock accuracy an agent has access to is a ratio between the average duration of the protocol and the uncertainty in their timing. This suggests that if we want to perform a given gate, i.e., fixed rotation angle $\theta$, with a timer whose timing uncertainty $\sigma$ is also fixed, carrying out a longer operation allows one to obtain a higher fidelity, which is reminiscent of the accuracy-resolution trade-off for clocks examined recently in Ref.~\cite{flo_accuracy_resolution}. At the cost of speed of computation one can obtain a higher average gate fidelity, provided one suitably adapts the intensity of the control parameter (e.g. Rabi frequency).

Our quantitative estimates indicate that, for experimentally relevant gate counts and durations, imperfect timekeeping may become a significant error source when the timing uncertainty $\sigma$ is on the picosecond scale or greater. It is crucial to emphasise that this is the uncertainty in timing the duration of each quantum gate, which must necessarily be very fast (e.g.~nanoseconds) in order to counteract environment-induced decoherence. Precise timing of such short time intervals remains an outstanding technical challenge. This should be contrasted with the sub-femtosecond uncertainty of atomic clocks, which is only achieved after integration times of seconds, hours or several days~\cite{Zheng2022} depending on the system.

From a foundational stance, our results have serious implications for the thermodynamics of quantum computation. Several papers explore this topic focusing only on the energetics of the system itself~\cite{auffeves,Deffner_2021,Stevens_2022,renner_chiribella,renner_chiribella_2}. While it is obvious that current technologies are dominated by the energy scales of the classical control (e.g.~refrigeration and laser pulses), there is hope in the community that---if larger circuits could be fit into those same fridges and near-reversible computation were possible---there would be an energetic advantage to quantum computation. By considering a single source of error we show that there is an intrinsic requirement that cannot be overcome and which scales with the circuit depth : the need for precise and accurate timekeeping for every gate. Yet mounting evidence underpinned by the discovery of thermodynamic uncertainty relations~\cite{Barato2015, Gingrich2016,Horowitz2020} demonstrates that precise timekeeping generally comes at a thermodynamic cost~\cite{thermal_clocks, Barato2016, ares, manu} (although see Ref.~\cite{Pietzonka2022} for a classical counterexample). For the autonomous thermal clocks considered in~\cite{thermal_clocks,manu,Barato2016}, the accuracy $N$ is shown to be bounded from above by
\begin{gather}
N \leq \frac{\Delta S_{\text{tick}}}{2}.
\label{entropy_bound}
\end{gather}
 where $\Delta S_{\text{tick}}$ is the entropy produced by the clock to generate each tick. Entropy production leads to heat dissipation into the environment, meaning that timekeeping implies a fundamental and inescapable contribution to the energetic cost of quantum computation which cannot be ignored.

It might seem that the message of this work is pessimistic, but the fact that timing is an issue in quantum computation is not unknown to those building physical devices. Algorithms have been developed in the field of optimal quantum control \cite{Rembold_2020,optimal_pulse_design,grapey} where \textit{bulk} errors can be mitigated by dynamically changing control parameters throughout the duration of a pulse. We expect that, by characterising the physics of specific sources of error within quantum computation as we have attempted in this work, new strategies to combat can be conceived.

\textit{Acknowledgements.---}The authors thank Pharnam (Faraj) Bakhshinezhad, Steve Campbell and Phila Rembold for insightful discussions. We especially thank Martin Ringbauer for sharing his knowledge of the state of the art of timing resolution in modern quantum control. J.X., P.E. and M.H. would like to acknowledge funding from the European Research Council (Consolidator grant `Cocoquest’ 101043705). F.M., P.E. M.T.M and M.H. further acknowledge funding by the European flagship on quantum technologies (`ASPECTS' consortium 101080167). P.E. and M.H. further acknowledge funds from the FQXi (FQXi-IAF19-03-S2) within the project ``Fueling quantum field machines with information''. M.T.M. is supported by a Royal Society-Science Foundation Ireland University Research Fellowship  (URF\textbackslash R1\textbackslash 221571).

\bibliography{ref}

\providecommand{\noopsort}[1]{}\providecommand{\singleletter}[1]{#1}%
\begin{thebibliography}{62}%
\makeatletter
\providecommand \@ifxundefined [1]{%
 \@ifx{#1\undefined}
}%
\providecommand \@ifnum [1]{%
 \ifnum #1\expandafter \@firstoftwo
 \else \expandafter \@secondoftwo
 \fi
}%
\providecommand \@ifx [1]{%
 \ifx #1\expandafter \@firstoftwo
 \else \expandafter \@secondoftwo
 \fi
}%
\providecommand \natexlab [1]{#1}%
\providecommand \enquote  [1]{``#1''}%
\providecommand \bibnamefont  [1]{#1}%
\providecommand \bibfnamefont [1]{#1}%
\providecommand \citenamefont [1]{#1}%
\providecommand \href@noop [0]{\@secondoftwo}%
\providecommand \href [0]{\begingroup \@sanitize@url \@href}%
\providecommand \@href[1]{\@@startlink{#1}\@@href}%
\providecommand \@@href[1]{\endgroup#1\@@endlink}%
\providecommand \@sanitize@url [0]{\catcode `\\12\catcode `\$12\catcode
  `\&12\catcode `\#12\catcode `\^12\catcode `\_12\catcode `\%12\relax}%
\providecommand \@@startlink[1]{}%
\providecommand \@@endlink[0]{}%
\providecommand \url  [0]{\begingroup\@sanitize@url \@url }%
\providecommand \@url [1]{\endgroup\@href {#1}{\urlprefix }}%
\providecommand \urlprefix  [0]{URL }%
\providecommand \Eprint [0]{\href }%
\providecommand \doibase [0]{http://dx.doi.org/}%
\providecommand \selectlanguage [0]{\@gobble}%
\providecommand \bibinfo  [0]{\@secondoftwo}%
\providecommand \bibfield  [0]{\@secondoftwo}%
\providecommand \translation [1]{[#1]}%
\providecommand \BibitemOpen [0]{}%
\providecommand \bibitemStop [0]{}%
\providecommand \bibitemNoStop [0]{.\EOS\space}%
\providecommand \EOS [0]{\spacefactor3000\relax}%
\providecommand \BibitemShut  [1]{\csname bibitem#1\endcsname}%
\let\auto@bib@innerbib\@empty
\bibitem [{\citenamefont {Wiseman}\ and\ \citenamefont
  {Milburn}(2009)}]{Wiseman2009}%
  \BibitemOpen
  \bibfield  {author} {\bibinfo {author} {\bibfnamefont {H.~M.}\ \bibnamefont
  {Wiseman}}\ and\ \bibinfo {author} {\bibfnamefont {G.~J.}\ \bibnamefont
  {Milburn}},\ }\href {\doibase 10.1017/CBO9780511813948} {\emph {\bibinfo
  {title} {Quantum {Measurement} and {Control}}}},\ \bibinfo {edition} {1st}\
  ed.\ (\bibinfo  {publisher} {Cambridge University Press},\ \bibinfo {year}
  {2009})\BibitemShut {NoStop}%
\bibitem [{\citenamefont {Cong}(2014)}]{Cong2014}%
  \BibitemOpen
  \bibfield  {author} {\bibinfo {author} {\bibfnamefont {S.}~\bibnamefont
  {Cong}},\ }\href
  {https://www.wiley.com/en-us/Control+of+Quantum+Systems:+Theory+and+Methods-p-9781118608128}
  {\emph {\bibinfo {title} {Control of quantum systems: theory and methods}}}\
  (\bibinfo  {publisher} {John Wiley \& Sons Inc},\ \bibinfo {address}
  {Singapore},\ \bibinfo {year} {2014})\BibitemShut {NoStop}%
\bibitem [{\citenamefont {Koch}\ \emph {et~al.}(2022)\citenamefont {Koch},
  \citenamefont {Boscain}, \citenamefont {Calarco}, \citenamefont {Dirr},
  \citenamefont {Filipp}, \citenamefont {Glaser}, \citenamefont {Kosloff},
  \citenamefont {Montangero}, \citenamefont {Schulte-Herbrüggen},
  \citenamefont {Sugny},\ and\ \citenamefont {Wilhelm}}]{Koch2022}%
  \BibitemOpen
  \bibfield  {author} {\bibinfo {author} {\bibfnamefont {C.~P.}\ \bibnamefont
  {Koch}}, \bibinfo {author} {\bibfnamefont {U.}~\bibnamefont {Boscain}},
  \bibinfo {author} {\bibfnamefont {T.}~\bibnamefont {Calarco}}, \bibinfo
  {author} {\bibfnamefont {G.}~\bibnamefont {Dirr}}, \bibinfo {author}
  {\bibfnamefont {S.}~\bibnamefont {Filipp}}, \bibinfo {author} {\bibfnamefont
  {S.~J.}\ \bibnamefont {Glaser}}, \bibinfo {author} {\bibfnamefont
  {R.}~\bibnamefont {Kosloff}}, \bibinfo {author} {\bibfnamefont
  {S.}~\bibnamefont {Montangero}}, \bibinfo {author} {\bibfnamefont
  {T.}~\bibnamefont {Schulte-Herbrüggen}}, \bibinfo {author} {\bibfnamefont
  {D.}~\bibnamefont {Sugny}}, \ and\ \bibinfo {author} {\bibfnamefont {F.~K.}\
  \bibnamefont {Wilhelm}},\ }\href {\doibase 10.1140/epjqt/s40507-022-00138-x}
  {\bibfield  {journal} {\bibinfo  {journal} {EPJ Quantum Technology}\ }\textbf
  {\bibinfo {volume} {9}},\ \bibinfo {pages} {19} (\bibinfo {year}
  {2022})}\BibitemShut {NoStop}%
\bibitem [{\citenamefont {Salecker}\ and\ \citenamefont
  {Wigner}(1958)}]{Salecker1958}%
  \BibitemOpen
  \bibfield  {author} {\bibinfo {author} {\bibfnamefont {H.}~\bibnamefont
  {Salecker}}\ and\ \bibinfo {author} {\bibfnamefont {E.~P.}\ \bibnamefont
  {Wigner}},\ }\href {\doibase 10.1103/PhysRev.109.571} {\bibfield  {journal}
  {\bibinfo  {journal} {Phys. Rev.}\ }\textbf {\bibinfo {volume} {109}},\
  \bibinfo {pages} {571} (\bibinfo {year} {1958})}\BibitemShut {NoStop}%
\bibitem [{\citenamefont {Peres}(1980)}]{peres}%
  \BibitemOpen
  \bibfield  {author} {\bibinfo {author} {\bibfnamefont {A.}~\bibnamefont
  {Peres}},\ }\href {\doibase 10.1119/1.12061} {\bibfield  {journal} {\bibinfo
  {journal} {American Journal of Physics}\ }\textbf {\bibinfo {volume} {48}},\
  \bibinfo {pages} {552} (\bibinfo {year} {1980})}\BibitemShut {NoStop}%
\bibitem [{\citenamefont {Erker}\ \emph {et~al.}(2017)\citenamefont {Erker},
  \citenamefont {Mitchison}, \citenamefont {Silva}, \citenamefont {Woods},
  \citenamefont {Brunner},\ and\ \citenamefont {Huber}}]{thermal_clocks}%
  \BibitemOpen
  \bibfield  {author} {\bibinfo {author} {\bibfnamefont {P.}~\bibnamefont
  {Erker}}, \bibinfo {author} {\bibfnamefont {M.~T.}\ \bibnamefont
  {Mitchison}}, \bibinfo {author} {\bibfnamefont {R.}~\bibnamefont {Silva}},
  \bibinfo {author} {\bibfnamefont {M.~P.}\ \bibnamefont {Woods}}, \bibinfo
  {author} {\bibfnamefont {N.}~\bibnamefont {Brunner}}, \ and\ \bibinfo
  {author} {\bibfnamefont {M.}~\bibnamefont {Huber}},\ }\href {\doibase
  10.1103/PhysRevX.7.031022} {\bibfield  {journal} {\bibinfo  {journal} {Phys.
  Rev. X}\ }\textbf {\bibinfo {volume} {7}},\ \bibinfo {pages} {031022}
  (\bibinfo {year} {2017})}\BibitemShut {NoStop}%
\bibitem [{\citenamefont {Woods}\ \emph {et~al.}(2019)\citenamefont {Woods},
  \citenamefont {Silva},\ and\ \citenamefont {Oppenheim}}]{Woods2019}%
  \BibitemOpen
  \bibfield  {author} {\bibinfo {author} {\bibfnamefont {M.~P.}\ \bibnamefont
  {Woods}}, \bibinfo {author} {\bibfnamefont {R.}~\bibnamefont {Silva}}, \ and\
  \bibinfo {author} {\bibfnamefont {J.}~\bibnamefont {Oppenheim}},\ }\href
  {\doibase 10.1007/s00023-018-0736-9} {\bibfield  {journal} {\bibinfo
  {journal} {Annales Henri Poincaré}\ }\textbf {\bibinfo {volume} {20}},\
  \bibinfo {pages} {125} (\bibinfo {year} {2019})}\BibitemShut {NoStop}%
\bibitem [{\citenamefont {Woods}(2021)}]{Woods2021autonomousticking}%
  \BibitemOpen
  \bibfield  {author} {\bibinfo {author} {\bibfnamefont {M.~P.}\ \bibnamefont
  {Woods}},\ }\href {\doibase 10.22331/q-2021-01-17-381} {\bibfield  {journal}
  {\bibinfo  {journal} {{Quantum}}\ }\textbf {\bibinfo {volume} {5}},\ \bibinfo
  {pages} {381} (\bibinfo {year} {2021})}\BibitemShut {NoStop}%
\bibitem [{\citenamefont {Schwarzhans}\ \emph {et~al.}(2021)\citenamefont
  {Schwarzhans}, \citenamefont {Lock}, \citenamefont {Erker}, \citenamefont
  {Friis},\ and\ \citenamefont {Huber}}]{manu}%
  \BibitemOpen
  \bibfield  {author} {\bibinfo {author} {\bibfnamefont {E.}~\bibnamefont
  {Schwarzhans}}, \bibinfo {author} {\bibfnamefont {M.~P.~E.}\ \bibnamefont
  {Lock}}, \bibinfo {author} {\bibfnamefont {P.}~\bibnamefont {Erker}},
  \bibinfo {author} {\bibfnamefont {N.}~\bibnamefont {Friis}}, \ and\ \bibinfo
  {author} {\bibfnamefont {M.}~\bibnamefont {Huber}},\ }\href {\doibase
  10.1103/PhysRevX.11.011046} {\bibfield  {journal} {\bibinfo  {journal} {Phys.
  Rev. X}\ }\textbf {\bibinfo {volume} {11}},\ \bibinfo {pages} {011046}
  (\bibinfo {year} {2021})}\BibitemShut {NoStop}%
\bibitem [{\citenamefont {Emerson}\ \emph {et~al.}(2005)\citenamefont
  {Emerson}, \citenamefont {Alicki},\ and\ \citenamefont
  {{\.{Z}}yczkowski}}]{Emerson_2005}%
  \BibitemOpen
  \bibfield  {author} {\bibinfo {author} {\bibfnamefont {J.}~\bibnamefont
  {Emerson}}, \bibinfo {author} {\bibfnamefont {R.}~\bibnamefont {Alicki}}, \
  and\ \bibinfo {author} {\bibfnamefont {K.}~\bibnamefont {{\.{Z}}yczkowski}},\
  }\href {\doibase 10.1088/1464-4266/7/10/021} {\bibfield  {journal} {\bibinfo
  {journal} {Journal of Optics B: Quantum and Semiclassical Optics}\ }\textbf
  {\bibinfo {volume} {7}},\ \bibinfo {pages} {S347} (\bibinfo {year}
  {2005})}\BibitemShut {NoStop}%
\bibitem [{\citenamefont {Magesan}\ \emph {et~al.}(2011)\citenamefont
  {Magesan}, \citenamefont {Gambetta},\ and\ \citenamefont
  {Emerson}}]{Magesan_2011}%
  \BibitemOpen
  \bibfield  {author} {\bibinfo {author} {\bibfnamefont {E.}~\bibnamefont
  {Magesan}}, \bibinfo {author} {\bibfnamefont {J.~M.}\ \bibnamefont
  {Gambetta}}, \ and\ \bibinfo {author} {\bibfnamefont {J.}~\bibnamefont
  {Emerson}},\ }\href {\doibase 10.1103/physrevlett.106.180504} {\bibfield
  {journal} {\bibinfo  {journal} {Physical Review Letters}\ }\textbf {\bibinfo
  {volume} {106}} (\bibinfo {year} {2011}),\
  10.1103/physrevlett.106.180504}\BibitemShut {NoStop}%
\bibitem [{\citenamefont {Carignan-Dugas}\ \emph {et~al.}(2018)\citenamefont
  {Carignan-Dugas}, \citenamefont {Boone}, \citenamefont {Wallman},\ and\
  \citenamefont {Emerson}}]{Carignan_Dugas_2018}%
  \BibitemOpen
  \bibfield  {author} {\bibinfo {author} {\bibfnamefont {A.}~\bibnamefont
  {Carignan-Dugas}}, \bibinfo {author} {\bibfnamefont {K.}~\bibnamefont
  {Boone}}, \bibinfo {author} {\bibfnamefont {J.~J.}\ \bibnamefont {Wallman}},
  \ and\ \bibinfo {author} {\bibfnamefont {J.}~\bibnamefont {Emerson}},\ }\href
  {\doibase 10.1088/1367-2630/aadcc7} {\bibfield  {journal} {\bibinfo
  {journal} {New Journal of Physics}\ }\textbf {\bibinfo {volume} {20}},\
  \bibinfo {pages} {092001} (\bibinfo {year} {2018})}\BibitemShut {NoStop}%
\bibitem [{\citenamefont {Ball}\ \emph {et~al.}(2016)\citenamefont {Ball},
  \citenamefont {Oliver},\ and\ \citenamefont {Biercuk}}]{Ball2016TheRO}%
  \BibitemOpen
  \bibfield  {author} {\bibinfo {author} {\bibfnamefont {H.}~\bibnamefont
  {Ball}}, \bibinfo {author} {\bibfnamefont {W.~D.}\ \bibnamefont {Oliver}}, \
  and\ \bibinfo {author} {\bibfnamefont {M.~J.}\ \bibnamefont {Biercuk}},\
  }\href {\doibase 10.1038/npjqi.2016.33} {\bibfield  {journal} {\bibinfo
  {journal} {npj Quantum Information}\ }\textbf {\bibinfo {volume} {2}}
  (\bibinfo {year} {2016}),\ 10.1038/npjqi.2016.33}\BibitemShut {NoStop}%
\bibitem [{\citenamefont {Schäfer}\ \emph {et~al.}(2018)\citenamefont
  {Schäfer}, \citenamefont {Ballance}, \citenamefont {Thirumalai},
  \citenamefont {Stephenson}, \citenamefont {Ballance}, \citenamefont
  {Steane},\ and\ \citenamefont {Lucas}}]{fast_quantum_gates}%
  \BibitemOpen
  \bibfield  {author} {\bibinfo {author} {\bibfnamefont {V.~M.}\ \bibnamefont
  {Schäfer}}, \bibinfo {author} {\bibfnamefont {C.~J.}\ \bibnamefont
  {Ballance}}, \bibinfo {author} {\bibfnamefont {K.}~\bibnamefont
  {Thirumalai}}, \bibinfo {author} {\bibfnamefont {L.~J.}\ \bibnamefont
  {Stephenson}}, \bibinfo {author} {\bibfnamefont {T.~G.}\ \bibnamefont
  {Ballance}}, \bibinfo {author} {\bibfnamefont {A.~M.}\ \bibnamefont
  {Steane}}, \ and\ \bibinfo {author} {\bibfnamefont {D.~M.}\ \bibnamefont
  {Lucas}},\ }\href {\doibase 10.1038/nature25737} {\bibfield  {journal}
  {\bibinfo  {journal} {Nature}\ }\textbf {\bibinfo {volume} {555}},\ \bibinfo
  {pages} {75–78} (\bibinfo {year} {2018})}\BibitemShut {NoStop}%
\bibitem [{\citenamefont {He}\ \emph {et~al.}(2022)\citenamefont {He},
  \citenamefont {Pakkiam}, \citenamefont {Gangat}, \citenamefont {Kewming},
  \citenamefont {Milburn},\ and\ \citenamefont {Fedorov}}]{he2022quantum}%
  \BibitemOpen
  \bibfield  {author} {\bibinfo {author} {\bibfnamefont {X.}~\bibnamefont
  {He}}, \bibinfo {author} {\bibfnamefont {P.}~\bibnamefont {Pakkiam}},
  \bibinfo {author} {\bibfnamefont {A.~A.}\ \bibnamefont {Gangat}}, \bibinfo
  {author} {\bibfnamefont {M.~J.}\ \bibnamefont {Kewming}}, \bibinfo {author}
  {\bibfnamefont {G.~J.}\ \bibnamefont {Milburn}}, \ and\ \bibinfo {author}
  {\bibfnamefont {A.}~\bibnamefont {Fedorov}},\ }\href@noop {} {\enquote
  {\bibinfo {title} {Quantum clock precision studied with a superconducting
  circuit},}\ } (\bibinfo {year} {2022}),\ \Eprint
  {http://arxiv.org/abs/2207.11043} {arXiv:2207.11043 [quant-ph]} \BibitemShut
  {NoStop}%
\bibitem [{\citenamefont {Jiang}\ \emph {et~al.}(2022)\citenamefont {Jiang},
  \citenamefont {Scott}, \citenamefont {Friesen},\ and\ \citenamefont
  {Saffman}}]{jiang_sensitivity_laser}%
  \BibitemOpen
  \bibfield  {author} {\bibinfo {author} {\bibfnamefont {X.}~\bibnamefont
  {Jiang}}, \bibinfo {author} {\bibfnamefont {J.}~\bibnamefont {Scott}},
  \bibinfo {author} {\bibfnamefont {M.}~\bibnamefont {Friesen}}, \ and\
  \bibinfo {author} {\bibfnamefont {M.}~\bibnamefont {Saffman}},\ }\href@noop
  {} {\enquote {\bibinfo {title} {Sensitivity of quantum gate fidelity to laser
  phase and intensity noise},}\ } (\bibinfo {year} {2022}),\ \Eprint
  {http://arxiv.org/abs/2210.11007} {arXiv:2210.11007 [quant-ph]} \BibitemShut
  {NoStop}%
\bibitem [{\citenamefont {Ballance}\ \emph {et~al.}(2016)\citenamefont
  {Ballance}, \citenamefont {Harty}, \citenamefont {Linke}, \citenamefont
  {Sepiol},\ and\ \citenamefont {Lucas}}]{ballance}%
  \BibitemOpen
  \bibfield  {author} {\bibinfo {author} {\bibfnamefont {C.~J.}\ \bibnamefont
  {Ballance}}, \bibinfo {author} {\bibfnamefont {T.~P.}\ \bibnamefont {Harty}},
  \bibinfo {author} {\bibfnamefont {N.~M.}\ \bibnamefont {Linke}}, \bibinfo
  {author} {\bibfnamefont {M.~A.}\ \bibnamefont {Sepiol}}, \ and\ \bibinfo
  {author} {\bibfnamefont {D.~M.}\ \bibnamefont {Lucas}},\ }\href {\doibase
  10.1103/PhysRevLett.117.060504} {\bibfield  {journal} {\bibinfo  {journal}
  {Phys. Rev. Lett.}\ }\textbf {\bibinfo {volume} {117}},\ \bibinfo {pages}
  {060504} (\bibinfo {year} {2016})}\BibitemShut {NoStop}%
\bibitem [{\citenamefont {Hrmo}\ \emph {et~al.}(2023)\citenamefont {Hrmo},
  \citenamefont {Wilhelm}, \citenamefont {Gerster}, \citenamefont {van Mourik},
  \citenamefont {Huber}, \citenamefont {Blatt}, \citenamefont {Schindler},
  \citenamefont {Monz},\ and\ \citenamefont {Ringbauer}}]{ringbauer_qudit}%
  \BibitemOpen
  \bibfield  {author} {\bibinfo {author} {\bibfnamefont {P.}~\bibnamefont
  {Hrmo}}, \bibinfo {author} {\bibfnamefont {B.}~\bibnamefont {Wilhelm}},
  \bibinfo {author} {\bibfnamefont {L.}~\bibnamefont {Gerster}}, \bibinfo
  {author} {\bibfnamefont {M.~W.}\ \bibnamefont {van Mourik}}, \bibinfo
  {author} {\bibfnamefont {M.}~\bibnamefont {Huber}}, \bibinfo {author}
  {\bibfnamefont {R.}~\bibnamefont {Blatt}}, \bibinfo {author} {\bibfnamefont
  {P.}~\bibnamefont {Schindler}}, \bibinfo {author} {\bibfnamefont
  {T.}~\bibnamefont {Monz}}, \ and\ \bibinfo {author} {\bibfnamefont
  {M.}~\bibnamefont {Ringbauer}},\ }\href {\doibase 10.1038/s41467-023-37375-2}
  {\bibfield  {journal} {\bibinfo  {journal} {Nature Communications}\ }\textbf
  {\bibinfo {volume} {14}} (\bibinfo {year} {2023}),\
  10.1038/s41467-023-37375-2}\BibitemShut {NoStop}%
\bibitem [{\citenamefont {Klenke}(2020)}]{Klenke2020}%
  \BibitemOpen
  \bibfield  {author} {\bibinfo {author} {\bibfnamefont {A.}~\bibnamefont
  {Klenke}},\ }\href {\doibase 10.1007/978-3-030-56402-5} {\emph {\bibinfo
  {title} {{P}robability {T}heory}}}\ (\bibinfo  {publisher} {Springer
  International Publishing},\ \bibinfo {year} {2020})\BibitemShut {NoStop}%
\bibitem [{foo()}]{footnote_fidelity}%
  \BibitemOpen
  \href@noop {} {}\bibinfo {note} {Note that here we are considering the
  fidelity of the \textit{entire} computation, implying that the algorithm is
  successful half the time. To achieve such a fidelity for an algorithm
  comprising many gates, the fidelity of individual gates would of course need
  to be much higher, i.e.~typically far above 99\%.}\BibitemShut {Stop}%
\bibitem [{\citenamefont {Breuer}\ and\ \citenamefont
  {Petruccione}(2007)}]{petruccione_breuer}%
  \BibitemOpen
  \bibfield  {author} {\bibinfo {author} {\bibfnamefont {H.-P.}\ \bibnamefont
  {Breuer}}\ and\ \bibinfo {author} {\bibfnamefont {F.}~\bibnamefont
  {Petruccione}},\ }\href {\doibase 10.1093/acprof:oso/9780199213900.001.0001}
  {\emph {\bibinfo {title} {{The Theory of Open Quantum Systems}}}}\ (\bibinfo
  {publisher} {Oxford University Press},\ \bibinfo {year} {2007})\BibitemShut
  {NoStop}%
\bibitem [{\citenamefont {Bylicka}\ \emph {et~al.}(2014)\citenamefont
  {Bylicka}, \citenamefont {Chru{\'{s}}ci{\'{n}}ski},\ and\ \citenamefont
  {Maniscalco}}]{Bylicka2014NonMarkovianityAR}%
  \BibitemOpen
  \bibfield  {author} {\bibinfo {author} {\bibfnamefont {B.}~\bibnamefont
  {Bylicka}}, \bibinfo {author} {\bibfnamefont {D.}~\bibnamefont
  {Chru{\'{s}}ci{\'{n}}ski}}, \ and\ \bibinfo {author} {\bibfnamefont
  {S.}~\bibnamefont {Maniscalco}},\ }\href {\doibase 10.1038/srep05720}
  {\bibfield  {journal} {\bibinfo  {journal} {Scientific Reports}\ }\textbf
  {\bibinfo {volume} {4}} (\bibinfo {year} {2014}),\
  10.1038/srep05720}\BibitemShut {NoStop}%
\bibitem [{\citenamefont {Nielsen}(2002)}]{Nielsen_2002}%
  \BibitemOpen
  \bibfield  {author} {\bibinfo {author} {\bibfnamefont {M.~A.}\ \bibnamefont
  {Nielsen}},\ }\href {\doibase 10.1016/s0375-9601(02)01272-0} {\bibfield
  {journal} {\bibinfo  {journal} {Physics Letters A}\ }\textbf {\bibinfo
  {volume} {303}},\ \bibinfo {pages} {249} (\bibinfo {year}
  {2002})}\BibitemShut {NoStop}%
\bibitem [{\citenamefont {Nielsen}\ and\ \citenamefont
  {Chuang}(2010)}]{nielsen_chuang_2010}%
  \BibitemOpen
  \bibfield  {author} {\bibinfo {author} {\bibfnamefont {M.~A.}\ \bibnamefont
  {Nielsen}}\ and\ \bibinfo {author} {\bibfnamefont {I.~L.}\ \bibnamefont
  {Chuang}},\ }\href {\doibase 10.1017/CBO9780511976667} {\emph {\bibinfo
  {title} {Quantum Computation and Quantum Information: 10th Anniversary
  Edition}}}\ (\bibinfo  {publisher} {Cambridge University Press},\ \bibinfo
  {year} {2010})\BibitemShut {NoStop}%
\bibitem [{\citenamefont {Amy}\ \emph {et~al.}(2018)\citenamefont {Amy},
  \citenamefont {Azimzadeh},\ and\ \citenamefont {Mosca}}]{Amy_2018}%
  \BibitemOpen
  \bibfield  {author} {\bibinfo {author} {\bibfnamefont {M.}~\bibnamefont
  {Amy}}, \bibinfo {author} {\bibfnamefont {P.}~\bibnamefont {Azimzadeh}}, \
  and\ \bibinfo {author} {\bibfnamefont {M.}~\bibnamefont {Mosca}},\ }\href
  {\doibase 10.1088/2058-9565/aad8ca} {\bibfield  {journal} {\bibinfo
  {journal} {Quantum Science and Technology}\ }\textbf {\bibinfo {volume}
  {4}},\ \bibinfo {pages} {015002} (\bibinfo {year} {2018})}\BibitemShut
  {NoStop}%
\bibitem [{\citenamefont {Rakyta}\ and\ \citenamefont
  {Zimbor{\'{a}}s}(2022)}]{CNOT_count_3}%
  \BibitemOpen
  \bibfield  {author} {\bibinfo {author} {\bibfnamefont {P.}~\bibnamefont
  {Rakyta}}\ and\ \bibinfo {author} {\bibfnamefont {Z.}~\bibnamefont
  {Zimbor{\'{a}}s}},\ }\href {\doibase 10.22331/q-2022-05-11-710} {\bibfield
  {journal} {\bibinfo  {journal} {{Quantum}}\ }\textbf {\bibinfo {volume}
  {6}},\ \bibinfo {pages} {710} (\bibinfo {year} {2022})}\BibitemShut {NoStop}%
\bibitem [{\citenamefont {Shende}\ \emph {et~al.}(2004)\citenamefont {Shende},
  \citenamefont {Markov},\ and\ \citenamefont {Bullock}}]{CNOT_count_4}%
  \BibitemOpen
  \bibfield  {author} {\bibinfo {author} {\bibfnamefont {V.~V.}\ \bibnamefont
  {Shende}}, \bibinfo {author} {\bibfnamefont {I.~L.}\ \bibnamefont {Markov}},
  \ and\ \bibinfo {author} {\bibfnamefont {S.~S.}\ \bibnamefont {Bullock}},\
  }\href {\doibase 10.1103/PhysRevA.69.062321} {\bibfield  {journal} {\bibinfo
  {journal} {Phys. Rev. A}\ }\textbf {\bibinfo {volume} {69}},\ \bibinfo
  {pages} {062321} (\bibinfo {year} {2004})}\BibitemShut {NoStop}%
\bibitem [{\citenamefont {Gheorghiu}\ \emph {et~al.}(2022)\citenamefont
  {Gheorghiu}, \citenamefont {Huang}, \citenamefont {Li}, \citenamefont
  {Mosca},\ and\ \citenamefont {Mukhopadhyay}}]{CNOT_count_2}%
  \BibitemOpen
  \bibfield  {author} {\bibinfo {author} {\bibfnamefont {V.}~\bibnamefont
  {Gheorghiu}}, \bibinfo {author} {\bibfnamefont {J.}~\bibnamefont {Huang}},
  \bibinfo {author} {\bibfnamefont {S.~M.}\ \bibnamefont {Li}}, \bibinfo
  {author} {\bibfnamefont {M.}~\bibnamefont {Mosca}}, \ and\ \bibinfo {author}
  {\bibfnamefont {P.}~\bibnamefont {Mukhopadhyay}},\ }\href {\doibase
  10.1109/TCAD.2022.3213210} {\bibfield  {journal} {\bibinfo  {journal} {IEEE
  Transactions on Computer-Aided Design of Integrated Circuits and Systems}\ ,\
  \bibinfo {pages} {1}} (\bibinfo {year} {2022})}\BibitemShut {NoStop}%
\bibitem [{\citenamefont {Carignan-Dugas}\ \emph
  {et~al.}(2019{\natexlab{a}})\citenamefont {Carignan-Dugas}, \citenamefont
  {Wallman},\ and\ \citenamefont {Emerson}}]{dugas_unitarity_bound}%
  \BibitemOpen
  \bibfield  {author} {\bibinfo {author} {\bibfnamefont {A.}~\bibnamefont
  {Carignan-Dugas}}, \bibinfo {author} {\bibfnamefont {J.~J.}\ \bibnamefont
  {Wallman}}, \ and\ \bibinfo {author} {\bibfnamefont {J.}~\bibnamefont
  {Emerson}},\ }\href {\doibase 10.1088/1367-2630/ab1800} {\bibfield  {journal}
  {\bibinfo  {journal} {New Journal of Physics}\ }\textbf {\bibinfo {volume}
  {21}},\ \bibinfo {pages} {053016} (\bibinfo {year}
  {2019}{\natexlab{a}})}\BibitemShut {NoStop}%
\bibitem [{\citenamefont {Carignan-Dugas}\ \emph
  {et~al.}(2019{\natexlab{b}})\citenamefont {Carignan-Dugas}, \citenamefont
  {Alexander},\ and\ \citenamefont {Emerson}}]{Carignan_Dugas_2019}%
  \BibitemOpen
  \bibfield  {author} {\bibinfo {author} {\bibfnamefont {A.}~\bibnamefont
  {Carignan-Dugas}}, \bibinfo {author} {\bibfnamefont {M.}~\bibnamefont
  {Alexander}}, \ and\ \bibinfo {author} {\bibfnamefont {J.}~\bibnamefont
  {Emerson}},\ }\href {\doibase 10.22331/q-2019-08-12-173} {\bibfield
  {journal} {\bibinfo  {journal} {Quantum}\ }\textbf {\bibinfo {volume} {3}},\
  \bibinfo {pages} {173} (\bibinfo {year} {2019}{\natexlab{b}})}\BibitemShut
  {NoStop}%
\bibitem [{\citenamefont {Boykin}\ \emph {et~al.}(1999)\citenamefont {Boykin},
  \citenamefont {Mor}, \citenamefont {Pulver}, \citenamefont {Roychowdhury},\
  and\ \citenamefont {Vatan}}]{boykin}%
  \BibitemOpen
  \bibfield  {author} {\bibinfo {author} {\bibfnamefont {P.}~\bibnamefont
  {Boykin}}, \bibinfo {author} {\bibfnamefont {T.}~\bibnamefont {Mor}},
  \bibinfo {author} {\bibfnamefont {M.}~\bibnamefont {Pulver}}, \bibinfo
  {author} {\bibfnamefont {V.}~\bibnamefont {Roychowdhury}}, \ and\ \bibinfo
  {author} {\bibfnamefont {F.}~\bibnamefont {Vatan}},\ }in\ \href {\doibase
  10.1109/SFFCS.1999.814621} {\emph {\bibinfo {booktitle} {40th Annual
  Symposium on Foundations of Computer Science (Cat. No.99CB37039)}}}\
  (\bibinfo {year} {1999})\ pp.\ \bibinfo {pages} {486--494}\BibitemShut
  {NoStop}%
\bibitem [{\citenamefont {Matsumoto}\ and\ \citenamefont
  {Amano}(2008)}]{matsumoto2008representation}%
  \BibitemOpen
  \bibfield  {author} {\bibinfo {author} {\bibfnamefont {K.}~\bibnamefont
  {Matsumoto}}\ and\ \bibinfo {author} {\bibfnamefont {K.}~\bibnamefont
  {Amano}},\ }\href@noop {} {\bibfield  {journal} {\bibinfo  {journal} {arXiv
  preprint arXiv:0806.3834}\ } (\bibinfo {year} {2008})}\BibitemShut {NoStop}%
\bibitem [{\citenamefont {Kasprowicz}\ \emph {et~al.}(2020)\citenamefont
  {Kasprowicz}, \citenamefont {Kulik}, \citenamefont {Gaska}, \citenamefont
  {Przywozki}, \citenamefont {Pozniak}, \citenamefont {Jarosinski},
  \citenamefont {Britton}, \citenamefont {Harty}, \citenamefont {Balance},
  \citenamefont {Zhang}, \citenamefont {Nadlinger}, \citenamefont {Slichter},
  \citenamefont {Allcock}, \citenamefont {Bourdeauducq}, \citenamefont
  {J\"{o}rdens},\ and\ \citenamefont {Pozniak}}]{artiq_sinara}%
  \BibitemOpen
  \bibfield  {author} {\bibinfo {author} {\bibfnamefont {G.}~\bibnamefont
  {Kasprowicz}}, \bibinfo {author} {\bibfnamefont {P.}~\bibnamefont {Kulik}},
  \bibinfo {author} {\bibfnamefont {M.}~\bibnamefont {Gaska}}, \bibinfo
  {author} {\bibfnamefont {T.}~\bibnamefont {Przywozki}}, \bibinfo {author}
  {\bibfnamefont {K.}~\bibnamefont {Pozniak}}, \bibinfo {author} {\bibfnamefont
  {J.}~\bibnamefont {Jarosinski}}, \bibinfo {author} {\bibfnamefont {J.~W.}\
  \bibnamefont {Britton}}, \bibinfo {author} {\bibfnamefont {T.}~\bibnamefont
  {Harty}}, \bibinfo {author} {\bibfnamefont {C.}~\bibnamefont {Balance}},
  \bibinfo {author} {\bibfnamefont {W.}~\bibnamefont {Zhang}}, \bibinfo
  {author} {\bibfnamefont {D.}~\bibnamefont {Nadlinger}}, \bibinfo {author}
  {\bibfnamefont {D.}~\bibnamefont {Slichter}}, \bibinfo {author}
  {\bibfnamefont {D.}~\bibnamefont {Allcock}}, \bibinfo {author} {\bibfnamefont
  {S.}~\bibnamefont {Bourdeauducq}}, \bibinfo {author} {\bibfnamefont
  {R.}~\bibnamefont {J\"{o}rdens}}, \ and\ \bibinfo {author} {\bibfnamefont
  {K.}~\bibnamefont {Pozniak}},\ }in\ \href {\doibase
  10.1364/QUANTUM.2020.QTu8B.14} {\emph {\bibinfo {booktitle} {OSA Quantum 2.0
  Conference}}}\ (\bibinfo  {publisher} {Optica Publishing Group},\ \bibinfo
  {year} {2020})\ p.\ \bibinfo {pages} {QTu8B.14}\BibitemShut {NoStop}%
\bibitem [{\citenamefont {Aharonov}\ and\ \citenamefont
  {Ben-Or}(1997)}]{aharonov1997fault}%
  \BibitemOpen
  \bibfield  {author} {\bibinfo {author} {\bibfnamefont {D.}~\bibnamefont
  {Aharonov}}\ and\ \bibinfo {author} {\bibfnamefont {M.}~\bibnamefont
  {Ben-Or}},\ }in\ \href {\doibase 10.1145/258533.258579} {\emph {\bibinfo
  {booktitle} {Proceedings of the twenty-ninth annual {ACM} symposium on Theory
  of computing - {STOC} {\textquotesingle}97}}}\ (\bibinfo  {publisher} {{ACM}
  Press},\ \bibinfo {year} {1997})\BibitemShut {NoStop}%
\bibitem [{\citenamefont {Bravyi}\ and\ \citenamefont
  {Kitaev}(2005)}]{bravyi_kitaev_fault}%
  \BibitemOpen
  \bibfield  {author} {\bibinfo {author} {\bibfnamefont {S.}~\bibnamefont
  {Bravyi}}\ and\ \bibinfo {author} {\bibfnamefont {A.}~\bibnamefont
  {Kitaev}},\ }\href {\doibase 10.1103/PhysRevA.71.022316} {\bibfield
  {journal} {\bibinfo  {journal} {Phys. Rev. A}\ }\textbf {\bibinfo {volume}
  {71}},\ \bibinfo {pages} {022316} (\bibinfo {year} {2005})}\BibitemShut
  {NoStop}%
\bibitem [{\citenamefont {Knill}\ \emph {et~al.}(1998)\citenamefont {Knill},
  \citenamefont {Laflamme},\ and\ \citenamefont {Zurek}}]{Knill_1998}%
  \BibitemOpen
  \bibfield  {author} {\bibinfo {author} {\bibfnamefont {E.}~\bibnamefont
  {Knill}}, \bibinfo {author} {\bibfnamefont {R.}~\bibnamefont {Laflamme}}, \
  and\ \bibinfo {author} {\bibfnamefont {W.~H.}\ \bibnamefont {Zurek}},\ }\href
  {\doibase 10.1098/rspa.1998.0166} {\bibfield  {journal} {\bibinfo  {journal}
  {Proceedings of the Royal Society of London. Series A: Mathematical, Physical
  and Engineering Sciences}\ }\textbf {\bibinfo {volume} {454}},\ \bibinfo
  {pages} {365} (\bibinfo {year} {1998})}\BibitemShut {NoStop}%
\bibitem [{\citenamefont {Malabarba}\ \emph {et~al.}(2015)\citenamefont
  {Malabarba}, \citenamefont {Short},\ and\ \citenamefont
  {Kammerlander}}]{Malabarba2015}%
  \BibitemOpen
  \bibfield  {author} {\bibinfo {author} {\bibfnamefont {A.~S.~L.}\
  \bibnamefont {Malabarba}}, \bibinfo {author} {\bibfnamefont {A.~J.}\
  \bibnamefont {Short}}, \ and\ \bibinfo {author} {\bibfnamefont
  {P.}~\bibnamefont {Kammerlander}},\ }\href {\doibase
  10.1088/1367-2630/17/4/045027} {\bibfield  {journal} {\bibinfo  {journal}
  {New Journal of Physics}\ }\textbf {\bibinfo {volume} {17}},\ \bibinfo
  {pages} {045027} (\bibinfo {year} {2015})}\BibitemShut {NoStop}%
\bibitem [{\citenamefont {Woods}\ and\ \citenamefont
  {Horodecki}(2019)}]{WoodsHorodecki2019}%
  \BibitemOpen
  \bibfield  {author} {\bibinfo {author} {\bibfnamefont {M.~P.}\ \bibnamefont
  {Woods}}\ and\ \bibinfo {author} {\bibfnamefont {M.}~\bibnamefont
  {Horodecki}},\ }\href@noop {} {\enquote {\bibinfo {title} {The resource
  theoretic paradigm of quantum thermodynamics with control},}\ } (\bibinfo
  {year} {2019}),\ \Eprint {http://arxiv.org/abs/1912.05562} {arXiv:1912.05562
  [quant-ph]} \BibitemShut {NoStop}%
\bibitem [{\citenamefont {Taranto}\ \emph {et~al.}(2023)\citenamefont
  {Taranto}, \citenamefont {Bakhshinezhad}, \citenamefont {Bluhm},
  \citenamefont {Silva}, \citenamefont {Friis}, \citenamefont {Lock},
  \citenamefont {Vitagliano}, \citenamefont {Binder}, \citenamefont {Debarba},
  \citenamefont {Schwarzhans}, \citenamefont {Clivaz},\ and\ \citenamefont
  {Huber}}]{Taranto2021}%
  \BibitemOpen
  \bibfield  {author} {\bibinfo {author} {\bibfnamefont {P.}~\bibnamefont
  {Taranto}}, \bibinfo {author} {\bibfnamefont {F.}~\bibnamefont
  {Bakhshinezhad}}, \bibinfo {author} {\bibfnamefont {A.}~\bibnamefont
  {Bluhm}}, \bibinfo {author} {\bibfnamefont {R.}~\bibnamefont {Silva}},
  \bibinfo {author} {\bibfnamefont {N.}~\bibnamefont {Friis}}, \bibinfo
  {author} {\bibfnamefont {M.~P.}\ \bibnamefont {Lock}}, \bibinfo {author}
  {\bibfnamefont {G.}~\bibnamefont {Vitagliano}}, \bibinfo {author}
  {\bibfnamefont {F.~C.}\ \bibnamefont {Binder}}, \bibinfo {author}
  {\bibfnamefont {T.}~\bibnamefont {Debarba}}, \bibinfo {author} {\bibfnamefont
  {E.}~\bibnamefont {Schwarzhans}}, \bibinfo {author} {\bibfnamefont
  {F.}~\bibnamefont {Clivaz}}, \ and\ \bibinfo {author} {\bibfnamefont
  {M.}~\bibnamefont {Huber}},\ }\href {\doibase 10.1103/PRXQuantum.4.010332}
  {\bibfield  {journal} {\bibinfo  {journal} {PRX Quantum}\ }\textbf {\bibinfo
  {volume} {4}},\ \bibinfo {pages} {010332} (\bibinfo {year}
  {2023})}\BibitemShut {NoStop}%
\bibitem [{\citenamefont {Park}\ \emph {et~al.}(2016)\citenamefont {Park},
  \citenamefont {Rodriguez-Briones}, \citenamefont {Feng}, \citenamefont
  {Rahimi}, \citenamefont {Baugh},\ and\ \citenamefont
  {Laflamme}}]{park2016heat}%
  \BibitemOpen
  \bibfield  {author} {\bibinfo {author} {\bibfnamefont {D.~K.}\ \bibnamefont
  {Park}}, \bibinfo {author} {\bibfnamefont {N.~A.}\ \bibnamefont
  {Rodriguez-Briones}}, \bibinfo {author} {\bibfnamefont {G.}~\bibnamefont
  {Feng}}, \bibinfo {author} {\bibfnamefont {R.}~\bibnamefont {Rahimi}},
  \bibinfo {author} {\bibfnamefont {J.}~\bibnamefont {Baugh}}, \ and\ \bibinfo
  {author} {\bibfnamefont {R.}~\bibnamefont {Laflamme}},\ }\href@noop {}
  {\bibfield  {journal} {\bibinfo  {journal} {Electron Spin Resonance (ESR)
  Based Quantum Computing}\ ,\ \bibinfo {pages} {227}} (\bibinfo {year}
  {2016})}\BibitemShut {NoStop}%
\bibitem [{\citenamefont {Bäumer}\ \emph {et~al.}(2019)\citenamefont
  {Bäumer}, \citenamefont {Perarnau-Llobet}, \citenamefont {Kammerlander},
  \citenamefont {Wilming},\ and\ \citenamefont
  {Renner}}]{imperfect_thermalisation}%
  \BibitemOpen
  \bibfield  {author} {\bibinfo {author} {\bibfnamefont {E.}~\bibnamefont
  {Bäumer}}, \bibinfo {author} {\bibfnamefont {M.}~\bibnamefont
  {Perarnau-Llobet}}, \bibinfo {author} {\bibfnamefont {P.}~\bibnamefont
  {Kammerlander}}, \bibinfo {author} {\bibfnamefont {H.}~\bibnamefont
  {Wilming}}, \ and\ \bibinfo {author} {\bibfnamefont {R.}~\bibnamefont
  {Renner}},\ }\href {\doibase 10.22331/q-2019-06-24-153} {\bibfield  {journal}
  {\bibinfo  {journal} {Quantum}\ }\textbf {\bibinfo {volume} {3}},\ \bibinfo
  {pages} {153} (\bibinfo {year} {2019})}\BibitemShut {NoStop}%
\bibitem [{\citenamefont {Clivaz}\ \emph
  {et~al.}(2019{\natexlab{a}})\citenamefont {Clivaz}, \citenamefont {Silva},
  \citenamefont {Haack}, \citenamefont {Brask}, \citenamefont {Brunner},\ and\
  \citenamefont {Huber}}]{Clivaz_2019_cost}%
  \BibitemOpen
  \bibfield  {author} {\bibinfo {author} {\bibfnamefont {F.}~\bibnamefont
  {Clivaz}}, \bibinfo {author} {\bibfnamefont {R.}~\bibnamefont {Silva}},
  \bibinfo {author} {\bibfnamefont {G.}~\bibnamefont {Haack}}, \bibinfo
  {author} {\bibfnamefont {J.~B.}\ \bibnamefont {Brask}}, \bibinfo {author}
  {\bibfnamefont {N.}~\bibnamefont {Brunner}}, \ and\ \bibinfo {author}
  {\bibfnamefont {M.}~\bibnamefont {Huber}},\ }\href {\doibase
  10.1103/physreve.100.042130} {\bibfield  {journal} {\bibinfo  {journal}
  {Physical Review E}\ }\textbf {\bibinfo {volume} {100}} (\bibinfo {year}
  {2019}{\natexlab{a}}),\ 10.1103/physreve.100.042130}\BibitemShut {NoStop}%
\bibitem [{\citenamefont {Clivaz}\ \emph
  {et~al.}(2019{\natexlab{b}})\citenamefont {Clivaz}, \citenamefont {Silva},
  \citenamefont {Haack}, \citenamefont {Brask}, \citenamefont {Brunner},\ and\
  \citenamefont {Huber}}]{Clivaz_2019_bound}%
  \BibitemOpen
  \bibfield  {author} {\bibinfo {author} {\bibfnamefont {F.}~\bibnamefont
  {Clivaz}}, \bibinfo {author} {\bibfnamefont {R.}~\bibnamefont {Silva}},
  \bibinfo {author} {\bibfnamefont {G.}~\bibnamefont {Haack}}, \bibinfo
  {author} {\bibfnamefont {J.~B.}\ \bibnamefont {Brask}}, \bibinfo {author}
  {\bibfnamefont {N.}~\bibnamefont {Brunner}}, \ and\ \bibinfo {author}
  {\bibfnamefont {M.}~\bibnamefont {Huber}},\ }\href {\doibase
  10.1103/physrevlett.123.170605} {\bibfield  {journal} {\bibinfo  {journal}
  {Physical Review Letters}\ }\textbf {\bibinfo {volume} {123}} (\bibinfo
  {year} {2019}{\natexlab{b}}),\ 10.1103/physrevlett.123.170605}\BibitemShut
  {NoStop}%
\bibitem [{\citenamefont {Silva}\ \emph {et~al.}(2016)\citenamefont {Silva},
  \citenamefont {Manzano}, \citenamefont {Skrzypczyk},\ and\ \citenamefont
  {Brunner}}]{Silva_2016}%
  \BibitemOpen
  \bibfield  {author} {\bibinfo {author} {\bibfnamefont {R.}~\bibnamefont
  {Silva}}, \bibinfo {author} {\bibfnamefont {G.}~\bibnamefont {Manzano}},
  \bibinfo {author} {\bibfnamefont {P.}~\bibnamefont {Skrzypczyk}}, \ and\
  \bibinfo {author} {\bibfnamefont {N.}~\bibnamefont {Brunner}},\ }\href
  {\doibase 10.1103/physreve.94.032120} {\bibfield  {journal} {\bibinfo
  {journal} {Physical Review E}\ }\textbf {\bibinfo {volume} {94}} (\bibinfo
  {year} {2016}),\ 10.1103/physreve.94.032120}\BibitemShut {NoStop}%
\bibitem [{\citenamefont {Xuereb}\ \emph {et~al.}(2023)\citenamefont {Xuereb},
  \citenamefont {Erker}, \citenamefont {Meier}, \citenamefont {Mitchison},\
  and\ \citenamefont {Huber}}]{sup_mat}%
  \BibitemOpen
  \bibfield  {author} {\bibinfo {author} {\bibfnamefont {J.}~\bibnamefont
  {Xuereb}}, \bibinfo {author} {\bibfnamefont {P.}~\bibnamefont {Erker}},
  \bibinfo {author} {\bibfnamefont {F.}~\bibnamefont {Meier}}, \bibinfo
  {author} {\bibfnamefont {M.~T.}\ \bibnamefont {Mitchison}}, \ and\ \bibinfo
  {author} {\bibfnamefont {M.}~\bibnamefont {Huber}},\ }\href {\doibase To be
  added by publisher} {\enquote {\bibinfo {title} {Supplemental material : The
  impact of imperfect timekeeping on quantum control},}\ } (\bibinfo {year}
  {2023})\BibitemShut {NoStop}%
\bibitem [{\citenamefont {Meier}\ \emph {et~al.}(2023)\citenamefont {Meier},
  \citenamefont {Schwarzhans}, \citenamefont {Erker},\ and\ \citenamefont
  {Huber}}]{flo_accuracy_resolution}%
  \BibitemOpen
  \bibfield  {author} {\bibinfo {author} {\bibfnamefont {F.}~\bibnamefont
  {Meier}}, \bibinfo {author} {\bibfnamefont {E.}~\bibnamefont {Schwarzhans}},
  \bibinfo {author} {\bibfnamefont {P.}~\bibnamefont {Erker}}, \ and\ \bibinfo
  {author} {\bibfnamefont {M.}~\bibnamefont {Huber}},\ }\href {\doibase
  10.48550/ARXIV.2301.05173} {\enquote {\bibinfo {title} {Fundamental
  accuracy-resolution trade-off for timekeeping devices},}\ } (\bibinfo {year}
  {2023})\BibitemShut {NoStop}%
\bibitem [{\citenamefont {Zheng}\ \emph {et~al.}(2022)\citenamefont {Zheng},
  \citenamefont {Dolde}, \citenamefont {Lochab}, \citenamefont {Merriman},
  \citenamefont {Li},\ and\ \citenamefont {Kolkowitz}}]{Zheng2022}%
  \BibitemOpen
  \bibfield  {author} {\bibinfo {author} {\bibfnamefont {X.}~\bibnamefont
  {Zheng}}, \bibinfo {author} {\bibfnamefont {J.}~\bibnamefont {Dolde}},
  \bibinfo {author} {\bibfnamefont {V.}~\bibnamefont {Lochab}}, \bibinfo
  {author} {\bibfnamefont {B.~N.}\ \bibnamefont {Merriman}}, \bibinfo {author}
  {\bibfnamefont {H.}~\bibnamefont {Li}}, \ and\ \bibinfo {author}
  {\bibfnamefont {S.}~\bibnamefont {Kolkowitz}},\ }\href {\doibase
  10.1038/s41586-021-04344-y} {\bibfield  {journal} {\bibinfo  {journal}
  {Nature}\ }\textbf {\bibinfo {volume} {602}},\ \bibinfo {pages} {425}
  (\bibinfo {year} {2022})}\BibitemShut {NoStop}%
\bibitem [{\citenamefont {Amoretti}(2021)}]{auffeves}%
  \BibitemOpen
  \bibfield  {author} {\bibinfo {author} {\bibfnamefont {M.}~\bibnamefont
  {Amoretti}},\ }\href {\doibase 10.22331/qv-2021-04-26-52} {\bibfield
  {journal} {\bibinfo  {journal} {Quantum Views}\ }\textbf {\bibinfo {volume}
  {5}},\ \bibinfo {pages} {52} (\bibinfo {year} {2021})}\BibitemShut {NoStop}%
\bibitem [{\citenamefont {Deffner}(2021)}]{Deffner_2021}%
  \BibitemOpen
  \bibfield  {author} {\bibinfo {author} {\bibfnamefont {S.}~\bibnamefont
  {Deffner}},\ }\href {\doibase 10.1209/0295-5075/134/40002} {\bibfield
  {journal} {\bibinfo  {journal} {Europhysics Letters}\ }\textbf {\bibinfo
  {volume} {134}},\ \bibinfo {pages} {40002} (\bibinfo {year}
  {2021})}\BibitemShut {NoStop}%
\bibitem [{\citenamefont {Stevens}\ \emph {et~al.}(2022)\citenamefont
  {Stevens}, \citenamefont {Szombati}, \citenamefont {Maffei}, \citenamefont
  {Elouard}, \citenamefont {Assouly}, \citenamefont {Cottet}, \citenamefont
  {Dassonneville}, \citenamefont {Ficheux}, \citenamefont {Zeppetzauer},
  \citenamefont {Bienfait}, \citenamefont {Jordan}, \citenamefont
  {Auff{\`{e}}ves},\ and\ \citenamefont {Huard}}]{Stevens_2022}%
  \BibitemOpen
  \bibfield  {author} {\bibinfo {author} {\bibfnamefont {J.}~\bibnamefont
  {Stevens}}, \bibinfo {author} {\bibfnamefont {D.}~\bibnamefont {Szombati}},
  \bibinfo {author} {\bibfnamefont {M.}~\bibnamefont {Maffei}}, \bibinfo
  {author} {\bibfnamefont {C.}~\bibnamefont {Elouard}}, \bibinfo {author}
  {\bibfnamefont {R.}~\bibnamefont {Assouly}}, \bibinfo {author} {\bibfnamefont
  {N.}~\bibnamefont {Cottet}}, \bibinfo {author} {\bibfnamefont
  {R.}~\bibnamefont {Dassonneville}}, \bibinfo {author} {\bibfnamefont
  {Q.}~\bibnamefont {Ficheux}}, \bibinfo {author} {\bibfnamefont
  {S.}~\bibnamefont {Zeppetzauer}}, \bibinfo {author} {\bibfnamefont
  {A.}~\bibnamefont {Bienfait}}, \bibinfo {author} {\bibfnamefont
  {A.}~\bibnamefont {Jordan}}, \bibinfo {author} {\bibfnamefont
  {A.}~\bibnamefont {Auff{\`{e}}ves}}, \ and\ \bibinfo {author} {\bibfnamefont
  {B.}~\bibnamefont {Huard}},\ }\href {\doibase 10.1103/physrevlett.129.110601}
  {\bibfield  {journal} {\bibinfo  {journal} {Physical Review Letters}\
  }\textbf {\bibinfo {volume} {129}} (\bibinfo {year} {2022}),\
  10.1103/physrevlett.129.110601}\BibitemShut {NoStop}%
\bibitem [{\citenamefont {Chiribella}\ \emph {et~al.}(2021)\citenamefont
  {Chiribella}, \citenamefont {Yang},\ and\ \citenamefont
  {Renner}}]{renner_chiribella}%
  \BibitemOpen
  \bibfield  {author} {\bibinfo {author} {\bibfnamefont {G.}~\bibnamefont
  {Chiribella}}, \bibinfo {author} {\bibfnamefont {Y.}~\bibnamefont {Yang}}, \
  and\ \bibinfo {author} {\bibfnamefont {R.}~\bibnamefont {Renner}},\ }\href
  {\doibase 10.1103/physrevx.11.021014} {\bibfield  {journal} {\bibinfo
  {journal} {Physical Review X}\ }\textbf {\bibinfo {volume} {11}} (\bibinfo
  {year} {2021}),\ 10.1103/physrevx.11.021014}\BibitemShut {NoStop}%
\bibitem [{\citenamefont {Chiribella}\ \emph {et~al.}(2022)\citenamefont
  {Chiribella}, \citenamefont {Meng}, \citenamefont {Renner},\ and\
  \citenamefont {Yung}}]{renner_chiribella_2}%
  \BibitemOpen
  \bibfield  {author} {\bibinfo {author} {\bibfnamefont {G.}~\bibnamefont
  {Chiribella}}, \bibinfo {author} {\bibfnamefont {F.}~\bibnamefont {Meng}},
  \bibinfo {author} {\bibfnamefont {R.}~\bibnamefont {Renner}}, \ and\ \bibinfo
  {author} {\bibfnamefont {M.-H.}\ \bibnamefont {Yung}},\ }\href {\doibase
  10.1038/s41467-022-34541-w} {\bibfield  {journal} {\bibinfo  {journal}
  {Nature Communications}\ }\textbf {\bibinfo {volume} {13}} (\bibinfo {year}
  {2022}),\ 10.1038/s41467-022-34541-w}\BibitemShut {NoStop}%
\bibitem [{\citenamefont {Barato}\ and\ \citenamefont
  {Seifert}(2015)}]{Barato2015}%
  \BibitemOpen
  \bibfield  {author} {\bibinfo {author} {\bibfnamefont {A.~C.}\ \bibnamefont
  {Barato}}\ and\ \bibinfo {author} {\bibfnamefont {U.}~\bibnamefont
  {Seifert}},\ }\href {\doibase 10.1103/PhysRevLett.114.158101} {\bibfield
  {journal} {\bibinfo  {journal} {Phys. Rev. Lett.}\ }\textbf {\bibinfo
  {volume} {114}},\ \bibinfo {pages} {158101} (\bibinfo {year}
  {2015})}\BibitemShut {NoStop}%
\bibitem [{\citenamefont {Gingrich}\ \emph {et~al.}(2016)\citenamefont
  {Gingrich}, \citenamefont {Horowitz}, \citenamefont {Perunov},\ and\
  \citenamefont {England}}]{Gingrich2016}%
  \BibitemOpen
  \bibfield  {author} {\bibinfo {author} {\bibfnamefont {T.~R.}\ \bibnamefont
  {Gingrich}}, \bibinfo {author} {\bibfnamefont {J.~M.}\ \bibnamefont
  {Horowitz}}, \bibinfo {author} {\bibfnamefont {N.}~\bibnamefont {Perunov}}, \
  and\ \bibinfo {author} {\bibfnamefont {J.~L.}\ \bibnamefont {England}},\
  }\href {\doibase 10.1103/PhysRevLett.116.120601} {\bibfield  {journal}
  {\bibinfo  {journal} {Phys. Rev. Lett.}\ }\textbf {\bibinfo {volume} {116}},\
  \bibinfo {pages} {120601} (\bibinfo {year} {2016})}\BibitemShut {NoStop}%
\bibitem [{\citenamefont {Horowitz}\ and\ \citenamefont
  {Gingrich}(2020)}]{Horowitz2020}%
  \BibitemOpen
  \bibfield  {author} {\bibinfo {author} {\bibfnamefont {J.~M.}\ \bibnamefont
  {Horowitz}}\ and\ \bibinfo {author} {\bibfnamefont {T.~R.}\ \bibnamefont
  {Gingrich}},\ }\href {\doibase 10.1038/s41567-019-0702-6} {\bibfield
  {journal} {\bibinfo  {journal} {Nature Physics}\ }\textbf {\bibinfo {volume}
  {16}},\ \bibinfo {pages} {15} (\bibinfo {year} {2020})}\BibitemShut {NoStop}%
\bibitem [{\citenamefont {Barato}\ and\ \citenamefont
  {Seifert}(2016)}]{Barato2016}%
  \BibitemOpen
  \bibfield  {author} {\bibinfo {author} {\bibfnamefont {A.~C.}\ \bibnamefont
  {Barato}}\ and\ \bibinfo {author} {\bibfnamefont {U.}~\bibnamefont
  {Seifert}},\ }\href {\doibase 10.1103/PhysRevX.6.041053} {\bibfield
  {journal} {\bibinfo  {journal} {Phys. Rev. X}\ }\textbf {\bibinfo {volume}
  {6}},\ \bibinfo {pages} {041053} (\bibinfo {year} {2016})}\BibitemShut
  {NoStop}%
\bibitem [{\citenamefont {Pearson}\ \emph {et~al.}(2021)\citenamefont
  {Pearson}, \citenamefont {Guryanova}, \citenamefont {Erker}, \citenamefont
  {Laird}, \citenamefont {Briggs}, \citenamefont {Huber},\ and\ \citenamefont
  {Ares}}]{ares}%
  \BibitemOpen
  \bibfield  {author} {\bibinfo {author} {\bibfnamefont {A.~N.}\ \bibnamefont
  {Pearson}}, \bibinfo {author} {\bibfnamefont {Y.}~\bibnamefont {Guryanova}},
  \bibinfo {author} {\bibfnamefont {P.}~\bibnamefont {Erker}}, \bibinfo
  {author} {\bibfnamefont {E.~A.}\ \bibnamefont {Laird}}, \bibinfo {author}
  {\bibfnamefont {G.~A.~D.}\ \bibnamefont {Briggs}}, \bibinfo {author}
  {\bibfnamefont {M.}~\bibnamefont {Huber}}, \ and\ \bibinfo {author}
  {\bibfnamefont {N.}~\bibnamefont {Ares}},\ }\href {\doibase
  10.1103/PhysRevX.11.021029} {\bibfield  {journal} {\bibinfo  {journal} {Phys.
  Rev. X}\ }\textbf {\bibinfo {volume} {11}},\ \bibinfo {pages} {021029}
  (\bibinfo {year} {2021})}\BibitemShut {NoStop}%
\bibitem [{\citenamefont {Pietzonka}(2022)}]{Pietzonka2022}%
  \BibitemOpen
  \bibfield  {author} {\bibinfo {author} {\bibfnamefont {P.}~\bibnamefont
  {Pietzonka}},\ }\href {\doibase 10.1103/PhysRevLett.128.130606} {\bibfield
  {journal} {\bibinfo  {journal} {Phys. Rev. Lett.}\ }\textbf {\bibinfo
  {volume} {128}},\ \bibinfo {pages} {130606} (\bibinfo {year}
  {2022})}\BibitemShut {NoStop}%
\bibitem [{\citenamefont {Rembold}\ \emph {et~al.}(2020)\citenamefont
  {Rembold}, \citenamefont {Oshnik}, \citenamefont {Müller}, \citenamefont
  {Montangero}, \citenamefont {Calarco},\ and\ \citenamefont
  {Neu}}]{Rembold_2020}%
  \BibitemOpen
  \bibfield  {author} {\bibinfo {author} {\bibfnamefont {P.}~\bibnamefont
  {Rembold}}, \bibinfo {author} {\bibfnamefont {N.}~\bibnamefont {Oshnik}},
  \bibinfo {author} {\bibfnamefont {M.~M.}\ \bibnamefont {Müller}}, \bibinfo
  {author} {\bibfnamefont {S.}~\bibnamefont {Montangero}}, \bibinfo {author}
  {\bibfnamefont {T.}~\bibnamefont {Calarco}}, \ and\ \bibinfo {author}
  {\bibfnamefont {E.}~\bibnamefont {Neu}},\ }\href {\doibase 10.1116/5.0006785}
  {\bibfield  {journal} {\bibinfo  {journal} {{AVS} Quantum Science}\ }\textbf
  {\bibinfo {volume} {2}},\ \bibinfo {pages} {024701} (\bibinfo {year}
  {2020})}\BibitemShut {NoStop}%
\bibitem [{\citenamefont {Li}\ \emph {et~al.}(2011)\citenamefont {Li},
  \citenamefont {Ruths}, \citenamefont {Yu}, \citenamefont {Arthanari},\ and\
  \citenamefont {Wagner}}]{optimal_pulse_design}%
  \BibitemOpen
  \bibfield  {author} {\bibinfo {author} {\bibfnamefont {J.-S.}\ \bibnamefont
  {Li}}, \bibinfo {author} {\bibfnamefont {J.}~\bibnamefont {Ruths}}, \bibinfo
  {author} {\bibfnamefont {T.-Y.}\ \bibnamefont {Yu}}, \bibinfo {author}
  {\bibfnamefont {H.}~\bibnamefont {Arthanari}}, \ and\ \bibinfo {author}
  {\bibfnamefont {G.}~\bibnamefont {Wagner}},\ }\href {\doibase
  10.1073/pnas.1009797108} {\bibfield  {journal} {\bibinfo  {journal}
  {Proceedings of the National Academy of Sciences}\ }\textbf {\bibinfo
  {volume} {108}},\ \bibinfo {pages} {1879} (\bibinfo {year}
  {2011})}\BibitemShut {NoStop}%
\bibitem [{\citenamefont {Machnes}\ \emph {et~al.}(2011)\citenamefont
  {Machnes}, \citenamefont {Sander}, \citenamefont {Glaser}, \citenamefont
  {de~Fouqui{\`{e}}res}, \citenamefont {Gruslys}, \citenamefont {Schirmer},\
  and\ \citenamefont {Schulte-Herbrüggen}}]{grapey}%
  \BibitemOpen
  \bibfield  {author} {\bibinfo {author} {\bibfnamefont {S.}~\bibnamefont
  {Machnes}}, \bibinfo {author} {\bibfnamefont {U.}~\bibnamefont {Sander}},
  \bibinfo {author} {\bibfnamefont {S.~J.}\ \bibnamefont {Glaser}}, \bibinfo
  {author} {\bibfnamefont {P.}~\bibnamefont {de~Fouqui{\`{e}}res}}, \bibinfo
  {author} {\bibfnamefont {A.}~\bibnamefont {Gruslys}}, \bibinfo {author}
  {\bibfnamefont {S.}~\bibnamefont {Schirmer}}, \ and\ \bibinfo {author}
  {\bibfnamefont {T.}~\bibnamefont {Schulte-Herbrüggen}},\ }\href {\doibase
  10.1103/physreva.84.022305} {\bibfield  {journal} {\bibinfo  {journal}
  {Physical Review A}\ }\textbf {\bibinfo {volume} {84}} (\bibinfo {year}
  {2011}),\ 10.1103/physreva.84.022305}\BibitemShut {NoStop}%
\bibitem [{\citenamefont {Wallman}\ \emph {et~al.}(2015)\citenamefont
  {Wallman}, \citenamefont {Granade}, \citenamefont {Harper},\ and\
  \citenamefont {Flammia}}]{unitarity_2015}%
  \BibitemOpen
  \bibfield  {author} {\bibinfo {author} {\bibfnamefont {J.}~\bibnamefont
  {Wallman}}, \bibinfo {author} {\bibfnamefont {C.}~\bibnamefont {Granade}},
  \bibinfo {author} {\bibfnamefont {R.}~\bibnamefont {Harper}}, \ and\ \bibinfo
  {author} {\bibfnamefont {S.~T.}\ \bibnamefont {Flammia}},\ }\href {\doibase
  10.1088/1367-2630/17/11/113020} {\bibfield  {journal} {\bibinfo  {journal}
  {New Journal of Physics}\ }\textbf {\bibinfo {volume} {17}},\ \bibinfo
  {pages} {113020} (\bibinfo {year} {2015})}\BibitemShut {NoStop}%
\end{thebibliography}%

\onecolumngrid
\appendix
\section*{Appendices}
\section{The Impact of Arbitrary Tick Distributions on Unitary Time Evolution}
\label{sec:arbitrary_tick}
Equation~\eqref{eq:gaussian_dephasing} is the specific expression for the time-evolution of the density matrix in case that the tick waiting time is Gaussian distributed.
While in a first approximation this may be justified, it is to be expected, that in general the waiting time distribution for a tick is not Gaussian \cite{manu,Woods2019}. Exponential decay for example is not Gaussian, but a primitive example for a thermal clock.
To generalize the dephasing result from eq.~\eqref{eq:gaussian_dephasing}, let us introduce a general tick probability density $p(t)$.
For the sake of the argument, let us assume that aside from being normalized, both the first $t_1$ and second moment $t_2$ of $p(t)$ exist. In the usual notation, we have
\begin{align}
    \tau \equiv t_1, \text{ and } \sigma^2\equiv t_2-t_1^2.
\end{align}
The matrix element $(m,n)$ of the density matrix $\rho$ after the unitary evolution (with unsharp timing) is given by an analogous expression as eq.~\eqref{eq:gaussian_dephasing},
\begin{align}
    \tilde{\rho}_{m,n}&=\int_{-\infty}^\infty dt p(t) e^{-i(E_m-E_n)t} \rho_{m,n}\\
    &=\rho_{m,n} e^{-i(E_m-E_n)\tau}\int_{-\infty}^\infty dt p(t-\tau) e^{-i(E_m-E_n)t}\label{eq:general_dephasing_integral}\\
    &=\rho_{m,n} e^{-i(E_m-E_n)\tau} \varphi_{T_\tau p}(E_m-E_n).\label{eq:general_dephasing_abbreviated}
\end{align}
From eq.~\eqref{eq:general_dephasing_integral} to eq.~\eqref{eq:general_dephasing_abbreviated}, we have abbreviated the Fourier transform of the shifted probability density function $T_\tau p (t):=p(t-\tau).$ The expression $\varphi_{T_\tau p}(\Omega)$ is known as the characteristic function of the probability density $T_\tau p$ \cite{Klenke2020}.
The contribution from the first moment of the tick probability density $p(t)$ has been factored out by means of a variable change. As a result, the unitary evolution factor $e^{-i\Omega t}$ is present in eq.~\eqref{eq:general_dephasing_abbreviated} and all the dephasing contributions are within the characteristic function $\varphi$.
In a next step we would like to give a more concrete characterization of the magnitude of the dephasing based on the uncertainty of the original tick distribution.
What we would expect is that a narrow tick time distribution (that is, $\sigma^2$ is vanishing) gives rise to little dephasing, as this case asymptotically coincides with perfect time-keeping (and thus unitary evolution).
Conversely, for a tick distribution with large uncertainty $\sigma$ in the time of arrival of the control clock's tick, we would expect stronger dephasing.
The strength of the dephasing is given by the magnitude of $\varphi(\Omega)$ (subscript implicit from now on). An additional complex phase in $\varphi(\Omega)$ is also possible if $p$ is not symmetric around $\tau$. The dephasing rate $\Gamma$ from the main text can be obtained as
\begin{align}
    \Gamma=-\log |\varphi(\Omega)|,
\end{align}
for a Rabi pulse inducing a frequency $\Omega$.
In the general case again, the two functions $\varphi(\Omega)$ and $p(t)$ are related by a Heisenberg-type uncertainty relation coming from the Fourier conjugation of the pair $(\Omega,t)$. One would expect a narrow distribution in time-domain of the tick probability density $p(t)$ to yield a wide distribution in energy-domain, that is $\varphi(\Omega)$ drops slowly. Consequently, for sharp tick distributions we would have little dephasing. The other way around, a wide distribution in time-domain gives a narrow energy-domain distribution, meaning $\varphi(\Omega)$ drops off quickly and the dephasing is relevant already for pulses addressing small transitions inducing a frequency $\Omega$.

\paragraph*{A counterexample.}
\begin{figure}
    \centering
    \includegraphics[width=\columnwidth]{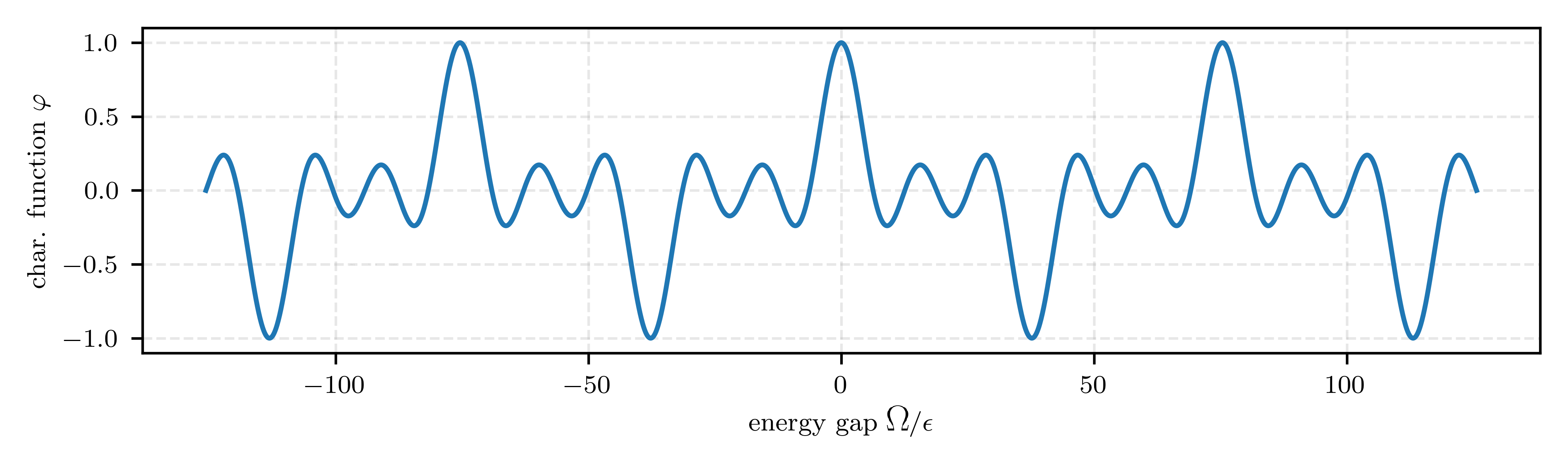}
    \caption{This plot shows the pathological characteristic function from eq.~\eqref{eq:pathological_char}. This example illustrates that there exist (non-Gaussian) tick probability densities whose dephasing rate $\Gamma$ does not diverge for large Rabi frequencies $\Omega$. The $x$-axis shows the values of $\Omega/\epsilon$ and the $y$-axis the value of $\varphi(\Omega)$ up to the global complex phase. It shows a dephasing function that periodically reaches unit magnitude, providing an example that even with \textbf{imperfect timekeeping}, that is $\sigma>0$, it is possible to have certain off-diagonal elements that \textbf{do not dephase} for given values of $E_n - E_m\neq 0$.}
    \label{fig:ugly_char_func}
\end{figure}
One may be tempted to conclude that a tick distribtion $p(t)$ with some variance $\sigma^2>0$ results in a lower bound for the amount of dephasing, given by $\varphi(\Omega)\leq g(\Omega,\sigma)<1.$ The term $g$ is some function that can be expressed in terms of the tick time uncertainty and energy gap $\Omega=E_m-E_n.$
There are pathological, comb-like tick probability distributions, for which such a statement is false. Take the distribution
\begin{align}
    p(t)=\frac{1}{n}\sum_{k=0}^{n-1}\delta\left(t-\frac{k}{n\epsilon}\right).
\end{align}
This distribution describes an imperfect clock in the sense that it has non-zero variance, $\sigma>0$. It describes a time-keeping device whose tick happens at multiples of $1/n\epsilon,$  with equal probability $1/n$. The characteristic function is given by the expression
\begin{align}
    \varphi(\Omega)&=\frac{1}{n}\sum_{k=0}^{n-1}e^{-i\frac{\Omega}{\epsilon}\frac{k}{n}}\\
    &=\frac{1}{n}\frac{1-e^{-i\frac{\Omega}{\epsilon}}}{1-e^{-i\frac{\Omega}{n\epsilon}}}\\
    &=\frac{e^{-i\frac{\Omega}{\epsilon}\left(\frac{1}{2}+\frac{1}{2n}\right)}}{n}\frac{\sin(\Omega/2\epsilon)}{\sin(\Omega/2n\epsilon)}.\label{eq:pathological_char}
\end{align}
Ignoring the global phase which comes from the fact that $p(t)$ does not have average $\tau=0,$ we obtain a function which periodically reaches unity when $\Omega$ is a multiple of $n\varepsilon$ (see Figure \ref{fig:ugly_char_func}).
\section{Fidelity Relationships}
\subsection{Average Channel Fidelity for a CNOT with imperfect timekeeping}
\label{sec:CNOT}
In the computational basis the CNOT gate can be expressed as 
\begin{gather}
\text{CNOT} = \ketbra{00}{00}+\ketbra{01}{01}+\ketbra{1+}{1+}-\ketbra{1-}{1-}
\end{gather}
and so can be expressed in terms of the matrix exponential of its generator as
\begin{gather}
\text{CNOT} = e^{-i\ketbra{1-}{1-}\pi}.
\end{gather}
Here, $\ketbra{1-}{1-}$ may be considered the Hamiltonian of the process carried out to enact CNOT with $\tau = \pi$ being its duration. This generator only acts on the $\ket{10}$, $\ket{11}$ subspace allowing us to consider this operation as one acting on an effective single qubit whose basis states are  $\ket{10}$ and $\ket{11}$. In this subspace, temporal error will affect the computation as follows
\begin{gather}
	\rho'_{\text{error}} = \int^{\infty}_{-\infty} \frac{1}{\sqrt{2\pi\sigma^2}}e^{\frac{(t-\pi)^2}{-2\sigma^2}} e^{-i\ketbra{1-}{1-}t}\rho e^{-i\ketbra{1-}{1-}t}  dt.
\end{gather}
Diagonalising $\ketbra{1-}{1-}$ in this effective one qubit subspace, we may determine the basis in which the dephasing will occur and so the Kraus operators corresponding to a time error channel for this operation. The states $\ket{+'} = \frac{\ket{10}+\ket{11}}{\sqrt{2}}$ and $\ket{-'} = \frac{\ket{10}-\ket{11}}{\sqrt{2}}$ diagonalise this Hamiltonian meaning that we can express an effective $\sigma_z$ operation on this subspace as
\begin{gather}
\sigma_z'= \ketbra{+'}{+'}-\ketbra{-'}{-'}.
\end{gather}
As a result we can apply the single qubit dephasing model from earlier to the subspace of this non-local qubit giving
\begin{align}
    K_1 = \sqrt{\frac{1 + e^{-\frac{\sigma^2\Omega^2}{2}}}{2}}\left(\ketbra{+'}{+'}+\ketbra{-'}{-'}\right) && K_2 =  \sqrt{\frac{1 - e^{-\frac{\sigma^2\Omega^2}{2}}}{2}}\left(\ketbra{+'}{+'}-\ketbra{-'}{-'}\right)
\end{align}
where $\Omega$ is the effective Rabi frequency induced by the control field on the the non-local qubit with basis states $\ket{+'}, \ket{-'}$, allowing us to obtain the average fidelity relation
\begin{gather}
	\overline{\mathcal{F}}(\mathcal{D}) = \frac{1}{3}\left(2 + e^{\frac{-\sigma^2\Omega^2}{2}}\right)
\end{gather}
which has the same form as eq.~\eqref{eq:infidelity}. Note that here we have Haar-averaged over this non-local subspace and not the total Hilbert space. Considering the total two qubit Hilbert space one has a dephasing channel described by the Kraus operators
\begin{align}
    K_1 = \sqrt{\frac{1 + e^{-\frac{\sigma^2\Omega^2}{2}}}{2}}I_1\oplus\left(\ketbra{+'}{+'}+\ketbra{-'}{-'}\right) && K_2 =  \sqrt{\frac{1 - e^{-\frac{\sigma^2\Omega^2}{2}}}{2}}I_1\oplus\left(\ketbra{+'}{+'}-\ketbra{-'}{-'}\right)
\label{eq:CNOT_kraus}
\end{align}
which gives the relation $\overline{\mathcal{F}}(\mathcal{D}) = \frac{1}{10}\left(7 + 3e^{\frac{-\sigma^2\Omega^2}{2}}\right)$, which is in general subtly larger than the previous quantity as the averaging now happening over a larger space.
\subsection{Imperfect Timekeeping on the SWAP Operation}
Similarly to CNOT, the SWAP operation can also be seen to act on a non-local single qubit subspace. In the Pauli Gate basis, the SWAP gate can be expressed as 
\begin{gather}
    \text{SWAP} = \frac{1}{2}\left(I\otimes I + X \otimes X + Y \otimes Y + Z \otimes Z \right)
\end{gather}
meaning that we may rewrite this as the matrix exponential 
\begin{gather}
\text{SWAP} = e^{-i\ketbra{\Psi^-}{\Psi^-}\pi}
\end{gather}
where $\ket{\Psi^-} = \frac{\ket{01} - \ket{10}}{\sqrt{2}}$ is one of the Bell basis states. As such the generator of the SWAP gate acts on the non-local virtual single qubit subspace spanned by $\ket{\Psi^+}$ and $\ket{\Psi^-}$. Therefore, in the same spirit as the derivation in the preceding section, imperfect timekeeping of the SWAP gate is given by a dephasing channel with Kraus operators
\begin{align}
    K_1 = \sqrt{\frac{1 + e^{-\frac{\sigma^2\Omega^2}{2}}}{2}}\left(\ketbra{\Psi^+}{\Psi^+}+\ketbra{\Psi^-}{\Psi^-}\right) && K_2 =  \sqrt{\frac{1 - e^{-\frac{\sigma^2\Omega^2}{2}}}{2}}\left(\ketbra{\Psi^+}{\Psi^+}-\ketbra{\Psi^-}{\Psi^-}\right)
\end{align}
where $\Omega$ is the Rabi frequency induced on the virtual qubit with basis states $\ket{\Psi^+}, \ket{\Psi^-}$.
\subsection{Impact of Concatenated Imperfect Timekeeping in a Quantum Circuit}
\label{sec:toffoli}
The nontrivial \textit{horizontal} scaling of the impact of imperfect timekeeping on average gate fidelity, is not straightforward to analyse in a generalised setting since the Kraus operators of the concatenation of $n$ dephasing channels are $2^n$ Kraus operators is given by the direct product $\left\{K_1,K_2\right\}^{\times \, n}$. This results in the average gate fidelity being computationally intractable in $n$ generally speaking. Despite this computational intractability, we can use the ideas from \cite{unitarity_2015,dugas_unitarity_bound,Carignan_Dugas_2019} to derive bounds on the average gate fidelity of a random circuit under the impact of concatenated gates, each suffering from imperfect timekeeping. In particular, consider a circuit on $n$ qubits with a circuit depth of $m$ and assume that at each time step $t$ in the circuit, a number $l_t \leq k$ CNOTs are applied to some $2l_t \leq n$ distinct qubits of the register where $L = \sum^m_{t=1} l_t$ is the total number of CNOTs in the circuit. Let us assume that each of the $L$ CNOTs is timed by an independent clock which with an identical Gaussian tick distribution having mean $\pi$ and variance $\sigma$. As a result of these tick distributions having a non-zero second moment one expects that at each time step $t$, $l_t$ non-local single qubit subspaces are dephased as in eq.\eqref{eq:CNOT_kraus}. This results in a global dephasing channel for each time step with Kraus operators from the set
\begin{align}
    \bm \Lambda_{(t)} := \left\{ K_{\alpha_1}\otimes \cdots \otimes  K_{\alpha_{l_t}}\otimes I_{n-2l_t}: \bm \alpha\in\{1,2\}^{l_t}\right\},
\end{align}
containing $2^{l_t}$ Kraus operators giving the dephasing of $l_t$ non-local single qubit subspaces corresponding to a different CNOT being applied in parallel in the $t$th time step. We denote the elements that live in this set $\bm \Lambda_{(t)}$ by $\Lambda^{\bm \alpha}_t\in \mathbf \Lambda_{(t)},$ for $ \bm \alpha\in\{1,2\}^{l_t}$ and 
\begin{align}
    \Lambda^{\bm\alpha}_t=K_{\alpha_1}\otimes \cdots \otimes  K_{\alpha_{l_t}}\otimes I_{n-2l_t},
\end{align}
with $K_1$ and $K_2$ given as calculated in eq.~\eqref{eq:CNOT_kraus}.
When it comes to calculating the products $\Lambda_t^{\bm \alpha \dagger} \Lambda_t^{\bm \alpha},$ the actual ordering of the entries of $\bm \alpha$ is irrelevant and we always have
\begin{align}
    \label{eq:LambdaDag_Lambda}
    \Lambda_t^{\bm\alpha \dagger} \Lambda_t^{\bm\alpha} &= \underbrace{\left(\frac{1+e^{-\Gamma}}{2}\right)^{l_t -i}\left(\frac{1-e^{-\Gamma}}{2}\right)^{i}}_{:=w_t^{\bm\alpha}} I_n,
\end{align}
where $i$ is the number of entries within $\bm\alpha$ that equals $1,$ i.e., $i=|\{0\leq k \leq l_t : \alpha_k = 1\}|.$
The prefactor $w_t^{\bm\alpha}$ therefore has a binomial multiplicity factor $\binom{l_t}{i},$ coming from the number of distinct choices $\alpha$ that have exactly $i$ entries equal to $1.$ For $m$ time steps we have a concatenated channel with Kraus operators from the set
\begin{align}
    \bm \Lambda = \bm \Lambda_{(m)} \cdot \bm \Lambda_{(m-1)} \cdots \bm \Lambda_{(1)},
\end{align}
which is the set defined through element-wise multiplication of the sets $\bm \Lambda_{(t)}.$ To be explicit, the set is given by the elements
\begin{align}
    \bm \Lambda = \left\{ \Lambda_m^{\bm \alpha_m}\Lambda_{m-1}^{\bm \alpha_{m-1}}\cdots \Lambda_1^{\bm \alpha_1} : \bm \alpha_m \in \{1,2\}^{l_m}, \bm \alpha_{m-1} \in \{1,2\}^{l_{m-1}}, \dots, \bm \alpha_1 \in \{1,2\}^{l_1} \right\},
\end{align}
and completely describes the dephasing coming from the concatenation of all the imperfectly timed CNOT gates. As a measure of how close this channel is to the desired unitary, we can compute the unitarity~\cite{unitarity_2015} of this channel. For a general noisy channel $\mathcal{M}$ with Kraus operators $K_i$ the unitarity is given by
\begin{gather}
 u(\mathcal{M}) = \frac{d^2 \Upsilon^2 (\mathcal{M}) - 1 }{d^2 - 1},
\end{gather}
where $\Upsilon = \sum_i \frac{||K_i||^2_2}{d}$, $||\cdot||_2$ is the Schatten 2-norm and $d$ is the dimension of the Kraus operator. This quantity is a measure of how much a noisy channel perturbs the purity preserving nature of a unitary it is applied to and is known to upper bound the average gate fidelity of concatenated channels~\cite{dugas_unitarity_bound,Carignan_Dugas_2019} as $\overline{\mathcal{F}} \leq \frac{d \Upsilon + 1}{d+1}$. For our global dephasing map we may calculate
\begin{align}
    \Upsilon^2 &= \sum_{\Lambda \in\bm\Lambda}\left(\frac{\|\Lambda\|_2^2}{2^n}\right)^2 \\
    &=\sum_{\bm\alpha_{m}\cdots \bm\alpha_1}  \frac{1}{2^{2n}}\left(\mathrm{tr}\left[\left(\Lambda_m^{\bm \alpha_m}\cdots\Lambda_1^{\bm \alpha_1} \right)^\dagger\left(\Lambda_m^{\bm \alpha_m}\cdots\Lambda_1^{\bm \alpha_1} \right)\right]\right)^2 \\
    &=\sum_{\bm\alpha_{m}\cdots \bm\alpha_1}  \frac{1}{2^{2n}}\left(\left(\prod_{t=1}^m w_t^{\bm\alpha_t}\right)\mathrm{tr}[I_n]\right)^2, \label{eq:IUD_2}
\end{align}
where we rewrote the trace as
\begin{align}
    \mathrm{tr}\left[\left(\Lambda_m^{\bm \alpha_m}\cdots\Lambda_1^{\bm \alpha_1} \right)^\dagger\left(\Lambda_m^{\bm \alpha_m}\cdots\Lambda_1^{\bm \alpha_1} \right)\right] &= \left(\prod_{t=1}^m w_t^{\bm\alpha_t}\right)\mathrm{tr}[I_n],
\end{align}
by using the property $\Lambda_t^{\bm \alpha_t \dagger}\Lambda_t^{\bm \alpha_t}= w_t^{\bm\alpha_t} I_{n},$ as outlined in eq.~\eqref{eq:LambdaDag_Lambda}.
Then, we simplify the sum over the multi-indices $\bm\alpha_m,\dots,\bm\alpha_1$ as a sum over how many entries of each multi-index $\bm\alpha_t$ equals $1,$ denoted by the number $i_t,$ and inserting a multiplicity factor of $\binom{l_t}{i_t}.$ Furthermore, we insert the definition of $w_t^{\bm\alpha_t}$ from eq.~\eqref{eq:LambdaDag_Lambda} and we obtain
\begin{align}
    \Upsilon^2&= \sum_{i_m\cdots i_1} \prod_{t=1}^m\binom{l_t}{i_t}\left(\left(\frac{1+e^{-\Gamma}}{2}\right)^{l_t-i_t}\left(\frac{1-e^{-\Gamma}}{2}\right)^{i_t}\right)^2 \label{eq:tr_resolving}\\
    &=\prod_{t=1}^m\left( \sum_{i_t=0}^{l_t} \binom{l_t}{i_t}\left(\frac{1+e^{-\Gamma}}{2}\right)^{2(l_t-i_t)}\left(\frac{1-e^{-\Gamma}}{2}\right)^{2 i_t}\right) \label{eq:product_sum_flip} \\
    &= \prod_{t=1}^m \left(\left(\frac{1+e^{-\Gamma}}{2}\right)^2+\left(\frac{1-e^{-\Gamma}}{2}\right)^2\right)^{l_t} \\
    &= \left(\frac{1 + e^{\frac{-\pi^2}{N}}}{2}\right)^{L},
\end{align}
by switching product and sum in eq.~\eqref{eq:product_sum_flip} according to $\sum_{i,j}a_i b_j = \left(\sum_i a_i\right)\left(\sum_j b_j\right)$.
This establishes the exact value for the unitarity of the global dephasing as a result of imperfectly timing $L$ CNOTs in an $m$ depth $n$ qubit circuit to be
\begin{gather}
u = \frac{2^{2n}\left(\frac{1 + e^{\frac{-\pi^2}{N}}}{2}\right)^{L}-1}{2^{2n}-1}.
\end{gather}
Using the results from~\cite{dugas_unitarity_bound,Carignan_Dugas_2019}, we find an upper bound on the average gate fidelity 
\begin{gather}
\overline{\mathcal{F}} \leq \frac{2^{n}\left(\frac{1 + e^{\frac{-\pi^2}{N}}}{2}\right)^{\frac{L}{2}}+1}{2^{n}+1}   
\end{gather}
where the average is a Haar average over the $n$ qubit initial input.
\section{An Imperfectly Timed Cooling Protocol}
\label{sec:cooling}
\begin{figure}[ht]
    \centering
    \includegraphics[width = 0.8\textwidth]{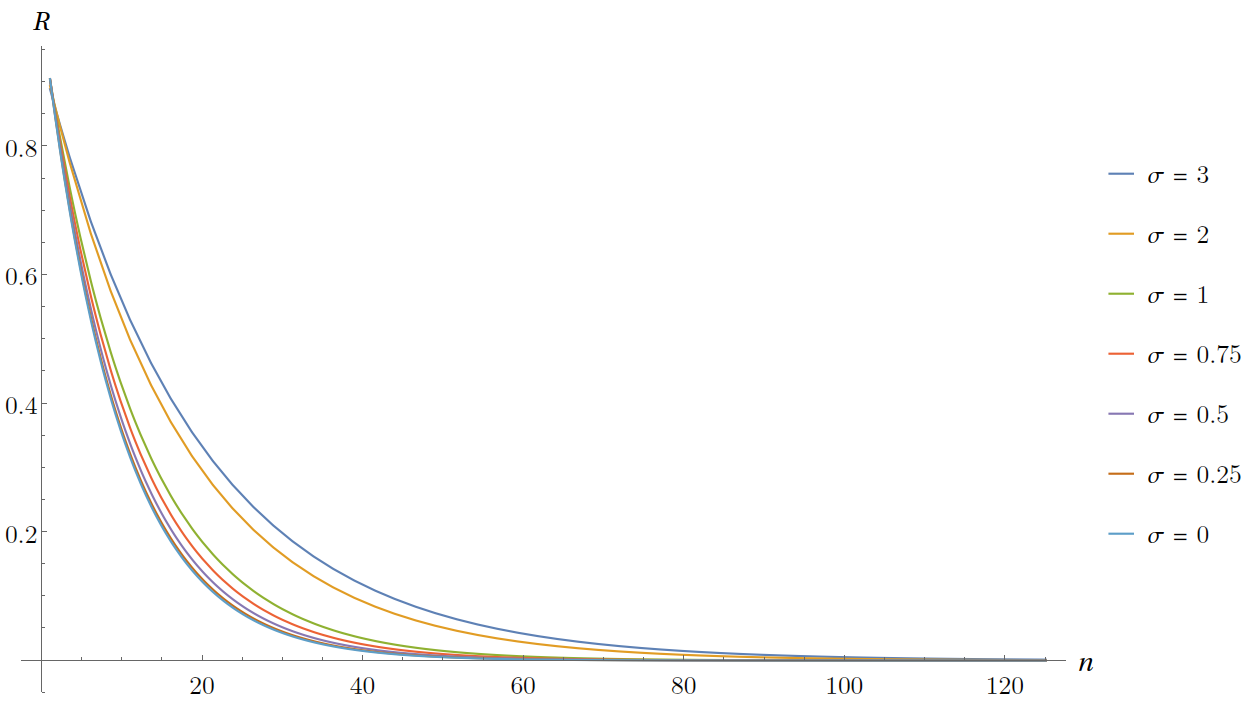}
    \caption{The rate of cooling, given by a central difference with step $h=0.1$, against the number of SWAPs carried out in the protocol for different imperfect timekeeping control expressed by $\sigma$ the variance of the tick distribution. In each plot the  virtual qubit occupation $P_v = 0.1$, target ground state population $r_v = 0.999$, initial qubit ground state population $r_s = 0.5$ were fixed, with only $\sigma$ varying.}
    \label{fig:rates}
\end{figure}
Assume that the physical qubit is in an initial thermal state $\rho_s$ at inverse temperature $\beta$ with Hamiltonian $H=-\omega_s \sigma_z$ giving
\begin{gather}
\rho_s = \frac{1+Z_s}{2}\ketbra{0}{0}_s + \frac{1-Z_s}{2}\ketbra{1}{1}_s 
\end{gather}
where $Z_s = \tanh(\beta \omega_s)$ is the partition function. The state of the machine is
\begin{gather}
 \rho_v = P_v\left(\frac{1+Z_v}{2}\ketbra{i}{i}_v + \frac{1-Z_v}{2}\ketbra{j}{j}\right) + \left(1 -P_V\right)\tilde{\rho}_v
\end{gather}
where $P_v = p_i + p_j < 1$ is the sum of the occupation of the energy levels of this ladder system which we are using to define our virtual qubit and $\tilde{\rho_v}$ is the state of the rest of the ladder system. Here the partition function is given by $Z_v = \tanh\left(\beta_v \Delta\right)$ where $\Delta = E_j - E_i$ and $\beta_v$ is the virtual qubit temperature.
Enacting the SWAP operation to cool the system results in 
\begin{multline}
\rho'=\text{SWAP}\rho_s \otimes \rho_v\text{SWAP} = \left(\frac{1+Z_s}{2}\right)P_v\left(\frac{1+Z_v}{2}\right) \ketbra{00}{00}_{sv} + \left(\frac{1-Z_s}{2}\right)P_v\left(\frac{1+Z_v}{2}\right) \ketbra{01}{01}_{sv}\\
+ \left(\frac{1+Z_s}{2}\right)P_v\left(\frac{1-Z_v}{2}\right) \ketbra{10}{10}_{sv} + \left(\frac{1-Z_s}{2}\right)P_v\left(\frac{1-Z_v}{2}\right) \ketbra{11}{11}_{sv} + \left(1 -P_V\right)\rho_s\otimes\tilde{\rho}_v
\end{multline}
and the reduced system state
\begin{gather}
\rho'_s = \left[P_v\left(\frac{1+Z_v}{2}\right) + \left(1 -P_V\right)\left(\frac{1+Z_s}{2}\right)\right]\ketbra{0}{0}_s + \left[P_v\left(\frac{1-Z_v}{2}\right) + \left(1 -P_V\right)\left(\frac{1-Z_s}{2}\right)\right]\ketbra{1}{1}_s.
\end{gather}
This implies that we may relate the normalised ground state probabilities as 
\begin{gather}
r'_s = P_vr_v + (1-P_v)r_s    
\end{gather}
that is with probability $P_v$ we have the ground state probability of the virtual qubit and with probability $(1-P_v)$ we have the initial ground state probability. With time error of tick variance $\sigma$ we instead have
\begin{align}
\rho'_{\text{error}} &= \mathcal{D}\left(\text{SWAP}\rho_s \otimes \rho_v\text{SWAP}\right)\\
&= \left(\frac{1+Z_s}{2}\right)P_v\left(\frac{1+Z_v}{2}\right) \ketbra{00}{00}_{sv} + \frac{P_v}{4}\left(1-Z_s Z_v + e^{-\frac{1}{2}\sigma^2{\Omega}^2}(Z_v - Z_s)\right) \ketbra{01}{01}_{sv}\\
&+\frac{P_v}{4}\left(1-Z_s Z_v - e^{-\frac{1}{2}\sigma^2\Omega^2}(Z_v - Z_s)\right) \ketbra{10}{10}_{sv} + \left(\frac{1-Z_s}{2}\right)P_v\left(\frac{1-Z_v}{2}\right) \ketbra{11}{11}_{sv} + \left(1 -P_V\right)\rho_s\otimes\tilde{\rho}_v
\end{align}
which can be expressed in terms of the target cooling state as 
\begin{gather}
\rho'_{\text{error}} = \rho' + P_v \frac{\left(1 -e^{-\sigma^2 \Omega^2/2}\right)}{2}\frac{\left(1+Z_s\right) - \left(1+Z_v\right)}{2}\ketbra{01}{01} + P_v \frac{\left(1 -e^{-\sigma^2\Omega^2/2}\right)}{2}\frac{\left(1-Z_s\right) - \left(1-Z_v\right)}{2}\ketbra{10}{10}
\end{gather}
and similarly the reduced state of the system being cooled
\begin{gather}
\rho'_{\text{error}_s} = \rho'_s + P_v \frac{\left(1 -e^{-\sigma^2 \omega^2/2}\right)}{2}\frac{\left(1+Z_s\right) - \left(1+Z_v\right)}{2}\ketbra{0}{0} + P_v \frac{\left(1 -e^{-\sigma^2 \Omega^2/2}\right)}{2}\frac{\left(1-Z_s\right) - \left(1-Z_v\right)}{2}\ketbra{1}{1}
\end{gather}
which allows us to obtain a relationship between normalised ground states as 
\begin{align}
r_{\text{error}} &= P_vr_v + (1-P_v)r_s + P_v\frac{\left(1 -e^{-\sigma^2 \Omega^2/2}\right)}{2}(r_s - r_v)\\
&= P_vr_v(1-p) + (1-P_v(1-p))r_s
\end{align}
where $p = \frac{\left(1 -e^{-\sigma^2 \Omega^2/2}\right)}{2}$, since $\theta = \Omega \tau$ and $N = (\frac{\tau}{\sigma})^2$ this can be recast as $p = \frac{\left(1 -e^{-\pi^2/2N}\right)}{2}$. In this way we obtain a relationship with the same form as that presented in \cite{Clivaz_2019_cost} with a modified probability $\widetilde{P_v} = P_v(1-p)$
\begin{gather}
    r_{\text{error}} = \widetilde{P_v}r_v + (1 - \widetilde{P_v})r_s
\end{gather}
this allows us to generalise to the application of $n$ repeated SWAP protocols with a refreshed virtual qubit at each step giving the recurrence relation
\begin{gather}
    \frac{r_v -r_{\text{error}}^{(n)}}{r_v - r_s} = (1 - \widetilde{P_v})^n
\end{gather}
which implies 
\begin{gather}
 r_{\text{error}}^{(n)} = r_v - (r_v - r_s)(1 -P_v(1-p))^n \label{eq:relation_error}
\end{gather}
and in the asymptotic regime of infinite swaps $n \to \infty$ we have that $r_{\text{error}} \to r_v$ and the system qubit is cooled down to the temperature of the virtual qubit despite error in timekeeping. 

In the case of perfect timekeeping, $\sigma = 0$ which gives $p = 0$ and recovers the relationship from \cite{Clivaz_2019_cost} whereas in the case that $\sigma = \infty$ we have $p = \frac{1}{2}$ in \eqref{eq:relation_error} meaning that whilst terrible timekeeping still gets you to the state you wish to cool to asymptotically, it does of course take longer. Physically, this may be interpreted as imperfect timekeeping giving you access to at worst half the virtual qubit population to swap within a given operation instead of the total virtual qubit population. Thus it is this partial swapping which impedes the cooling, but in some sense something is always swapped in the right direction provided that the virtual qubit's virtual temperature is lower than that of the system qubit to begin with.

Whilst we are still able to cool a qubit to an arbitrary target temperature with an imperfect clock controlling our cooling protocol it is clear that one would require more resources to do so. To explore this, a rate of cooling within this context can be obtained by taking the central difference of the recurrence relation eq.~\eqref{eq:relation_error}.
\begin{gather}
    R_h = \frac{r_{\text{error}}^{(n + \frac{h}{2})} -  r_{\text{error}}^{(n - \frac{h}{2})}}{h}
\end{gather}
for different values of $p$ given by the variance of the tick distribution. Here $h$ is the step of the finite difference. From eq.~\eqref{eq:relation_error} one finds the rate
\begin{gather}
    R_h = \left(\frac{r_s-r_v}{h}\right)\left[((1-\widetilde{P_v})^h - 1)(1-\widetilde{P_v})^{n-\frac{h}{2}}\right]
\end{gather}
which we plot in Fig~\ref{fig:rates} for fixed example values of $r_s, r_v$ and $P_v$ whilst varying $\sigma$. Showing that imperfect clocks require more SWAPs to cool to the target state.
\end{document}